\documentclass[12pt]{article}
\usepackage[a4paper,left=2cm,right=2cm,top=2.5cm,bottom=2.5cm]{geometry}
\onecolumn 

\usepackage{bbm}        
\usepackage{tikz}       
\usepackage{eurosym}    
\usepackage{graphicx}   
\usepackage{booktabs}   
\usepackage{xcolor}     
\usepackage{hyperref}   
\usepackage{float}      
\usepackage{amsthm}
\usepackage[section]{placeins} 
\usepackage{authblk}
\usepackage{hyperref}
\usepackage{orcidlink}
\usepackage{amsmath}
\usepackage[authoryear]{natbib}
\graphicspath{{Fig/}}
\usepackage{amssymb}

\title{A Dynamic Latent Space Model for Healthcare Mobility Networks: the Italian National Health Service case}

\author[1]{Cecilia Manente\,\orcidlink{0009-0007-2032-9208}\thanks{Corresponding author: \href{mailto:cecilia.manente@uniroma1.it}{cecilia.manente@uniroma1.it}}}
\author[1]{Marco Alfò\,\orcidlink{0000-0001-7651-6052}}
\author[2]{Silvia D'Angelo\,\orcidlink{0000-0003-3541-6797}}

\affil[1]{School of Statistical Sciences, Sapienza Università di Roma, Piazzale Aldo Moro 5, 00185 Rome, Italy}

\affil[2]{School of Computer Science and Statistics, Trinity College Dublin, Dublin 2, D02 PN40, Ireland}

\date{}

\begin{document}

\maketitle

\abstract{Healthcare mobility --- patients seeking treatment outside their territory of residence --- represents a major source of inequality and financial imbalance in decentralised health systems. In Italy, persistent north--south asymmetries in patient flows among Local Health Authorities (ASLs) have reinforced existing disparities within the National Health Service; yet the structural organisation and temporal dynamics of these flows remain poorly understood at the sub-regional level. We propose a Bayesian dynamic latent space model for directed weighted networks with a hurdle negative binomial likelihood, and apply it to administrative discharge records on mobility for hip replacement procedures among 109 Italian ASLs over 2018--2024. The model jointly addresses excess zeros, overdispersion and network dependence, while capturing directional heterogeneity through 
multiplicative sender and receiver effects and controlling for differences in territorial size via an appropriate exposure term. Applied to Italian mobility data, the model reveals the evolving geometry of the healthcare system, quantifies the disruption induced by the COVID-19 pandemic, and uncovers structural asymmetries in outward propensity and ASLs attractiveness. The framework provides a flexible tool for the statistical analysis of dynamic healthcare mobility networks with direct relevance to the monitoring and evaluation of territorial healthcare provision.}

\maketitle

\section{Introduction}
\label{introduction}

In recent decades, several countries have implemented decentralisation reforms in their national health services (NHSs), delegating varying degrees of responsibility for healthcare organisation and delivery to subnational governments \citep{adolph2012allocation}. In such contexts, the interplay between territorial autonomy and patients’ freedom to choose their providers results in cross-border utilisation of services, commonly referred to as \emph{healthcare mobility} \citep{balia2018interregional}. While all individuals necessarily travel some distance to access care, the term is generally applied more specifically to describe patients who seek treatment outside their territorial jurisdiction of residence \citep{zocchetti2012mobilita}.

Italy provides a particularly informative case. The National Health Service (\emph{Servizio Sanitario Nazionale}, SSN), established by Law~833/1978 as a universal system primarily funded through general taxation, integrates nationally defined principles and benefits with decentralised service delivery. A nationally guaranteed set of essential services (\emph{Livelli Essenziali di Assistenza}, LEA) is specified at central level, but the organisation and provision of care are largely delegated to 21 autonomous Regional Health Services, which may adapt hospital capacity, technological investment and service networks to local requirements \citep{balia2018interregional, moscone2012social}. Within this structure, citizens are enrolled with Local Health Authorities (\emph{Aziende Sanitarie Locali}, ASL) based on their residence, but retain the right to access SSN-funded care in public hospitals and accredited private facilities nationwide \citep{EuroObsItaly2024}. Hospital admissions outside the region of enrolment are reimbursed through interregional compensation schemes using diagnosis-related group (DRG) tariffs, resulting in explicit financial flows between regions \citep{neri2015interregional}. This institutional configuration—combining decentralisation, portability of entitlements, and free choice—results in substantial and persistent levels of interregional patient mobility. For example, in the period 2018–2019, approximately 680,000 hospital admissions involved patients treated outside their region of residence \citep{ISS2019}.

The geography of these flows often reveals a pronounced north–south asymmetry. Central and northern regions are typically net importers of patients and recipients of mobility-related financial resources, while southern regions tend to be net exporters of patients and experience corresponding financial outflows \citep{nante2021inter, carnazza2025monetary}. Economic compensation for net patient flows has provided additional resources to higher-performing central and northern regions, thereby reinforcing the existing north–south gradient within the Italian NHS \citep{carnazza2025monetary}. Despite decades of decentralisation and interregional competition, mobility flows remain substantial, indicating that quality differentials and structural imbalances persist. Thus, healthcare mobility reflects not only individual choice and clinical specialisation, but also territorial inequities within a decentralised universal health system \citep{moscone2012social, balia2018interregional, lu2025patient}.

Most empirical studies and institutional monitoring frameworks, such as the periodic reports by the National Agency for Regional Health Services (AGENAS), use the \emph{region} as the unit of analysis, treating each of the 21 Regional Health Services as a single node within the mobility network \citep{AGENAS2024, balia2018interregional, carnazza2025monetary}. While this aggregation aligns with financial compensation mechanisms, it obscures significant within-region heterogeneity. Italian regions vary considerably in population size and internal organisation, and individual ASLs within the same region may display distinct escape and attraction profiles based on proximity to regional borders, local hospital infrastructure, and the presence of centres of excellence in neighbouring jurisdictions \citep{balia2018interregional, moscone2012social}. Analysing mobility at the ASL level, rather than at the regional level, maintains the spatial granularity relevant to patient decision-making and reveals a more nuanced relational structure, including local hubs, peripheries, and community patterns. Furthermore, ASL-level flows can be aggregated to the regional level, ensuring comparability with existing regional analyses.

The mobility network capturing inter-regional patient movements constitutes a directed weighted network, where nodes represent territorial units and each directed edge indicates a non-negative count of hospital admissions by residents of the origin node in facilities of the destination node. Traditional quantitative studies of patient mobility have primarily employed gravity-type models, which relate bilateral flows to characteristics of the origin and destination, as well as a distance-decay term, typically assuming dyadwise conditional independence \citep{congdon2001development, balia2016spatial}. Although these models help explain supply and demand determinants, they consider each origin–destination pair independently and therefore cannot capture structural network features such as reciprocity, sender and receiver heterogeneity, transitivity, higher-order clustering, or the propagation of referral patterns and care-seeking behaviour throughout the system.

Network-based statistical models provide a natural alternative. Exponential random graph models (ERGMs) provide a principled network-based framework in which the full joint distribution of the adjacency matrix is parameterised using local network statistics, thereby accommodating dependencies across dyads \citep{robins2007introduction}. Dynamic extensions, such as the Separable Temporal ERGM proposed by \citet{krivitsky2014separable}, model tie formation and dissolution over time. However, ERGMs were initially developed for binary networks and are most commonly applied in that context; recently, extensions to valued or count-valued edges have been developed, but they can be computationally intensive in longitudinal settings where edge weights convey substantive information on flow intensity. Latent space models (LSMs) provide an alternative framework, particularly suited to weighted network data. Introduced by \citet{hoff2002latent}, the latent space approach assumes that each node occupies a position in a low-dimensional Euclidean space, with the probability of a tie decreasing as the latent distance between nodes increases. This construction induces transitivity and homophily, yields an interpretable geometric representation of network structure, and enables quantification of statistical uncertainty in node positions through posterior inference. The framework was extended by \citet{hoff2005bilinear} to directed and weighted relational data via bilinear multiplicative effects, and by \citet{handcock2007model} to model-based clustering of latent positions. 

Recent work on directed networks by \citet{lu2026mixed} relaxes the inherently symmetric geometric structure typically assumed in latent space models, allowing outgoing and incoming edges to follow distinct statistical distributions. In this framework, nodes are assigned separate representations in the latent space according to their roles as senders and receivers.
Here, we pursue a different but complementary objective. 
Rather than embedding directional asymmetry within the geometry of the latent space, we model the asymmetric component of inter-ASL flows through explicit sender and receiver parameters, estimated separately from the latent positions.
The latent space is thus reserved for capturing the symmetric, aggregate structure of patient exchange---the underlying geography of healthcare accessibility and referral patterns shared by both directions of flow---while directional heterogeneity is accounted for by the multiplicative effects. 
This separation allows for a more direct interpretation of the latent geometry as a map of structural proximity among ASLs, independently of their individual propensities to send or receive patients.

Longitudinal extensions represent the trajectory of each node through latent space as a smooth function of time, capturing the temporal evolution of the network while preserving interpretability \citep{sewell2015latent, DuranteDunson2016}. Although these methods have been applied in social, political and financial network analysis, their application to healthcare mobility networks remains, to the best of our knowledge, unexplored.

Another characteristic of mobility networks, particularly at fine spatial resolutions such as the ASL level, is the high prevalence of zero flows: many pairs of territorial units record no patient exchanges within a given year. Standard Poisson or negative binomial models conflate structural zeros—pairs with no underlying propensity to exchange patients—with sampling zeros that result from small but positive expected counts. Hurdle models, which decompose the likelihood into a binary component for the zero/positive outcome and a truncated count component for positive outcomes \citep{mullahy1986specification}, provide a more suitable framework by allowing these two processes to be governed by different sets of predictors and latent structures.

This paper integrates the aforementioned elements within a unified Bayesian framework. Italian health mobility is modelled as a dynamic directed weighted network among ASLs, observed over multiple years, using a dynamic latent space model with a hurdle negative binomial specification. By embedding ASL nodes in a low-dimensional latent space with time-varying positions, the model captures the evolving geometry of the Italian healthcare system, accounts for excess zeros and overdispersion, and yields an interpretable representation of structural asymmetries and their temporal dynamics. The rest of the paper is organised as follows. 
Section~\ref{sec:data} describes the administrative data source, the hip replacement cohort, and the structural properties of the inter-ASL origin--destination matrices.
Section~\ref{sec:model} introduces the dynamic latent space model with hurdle negative binomial likelihood and details the estimation algorithm. 
Section~\ref{sec:simulation} presents the simulation study assessing the performance of the proposed MCMC procedure. Section~\ref{sec:application} applies the framework to Italian ASL mobility data, comparing model specifications and analysing the estimated latent space, sender and receiver effects, and overdispersion dynamics. 
Section~\ref{sec:discussion} concludes with a discussion of the main findings, limitations, and directions for future research.

\section{Hip replacement patient mobility in the Italian SSN}
\label{sec:data}

\subsection{Administrative data source}
This  analysis uses administrative hospital discharge records (\emph{Schede di Dimissione Ospedaliera}, SDO) collected nationwide by the Italian Ministry of Health. Established in the early 1990s, the SDO system is periodically updated to ensure harmonised coding under the \emph{International Classification of Diseases, Ninth Revision, Clinical Modification} (ICD-9-CM) and consistency in reimbursement rules \citep{epicentro_sdo,balia2018interregional,AGENAS2024Mobility,carnazza2025monetary}. 
The system captures all SSN-funded inpatient discharges from both public hospitals and accredited private providers. As interregional financial compensation is based on these records, coverage of publicly reimbursed hospitalisations is effectively complete \citep{AGENAS2024Mobility}. The SDO database constitutes the official source for monitoring patient mobility and assessing performance within the SSN. Each record refers to a single inpatient episode, and it includes detailed information on the patient’s Local Health Authority (ASL) of residence, the ASL of the treating hospital, year of admission, assigned Diagnosis-Related Group (DRG), demographic characteristics (age and sex) and coded diagnoses and procedures \citep{MinSal_SDO}. These data underpin the \emph{Programma Nazionale Esiti} (PNE), the national programme for evaluating hospital performance and quality of care within the SSN \citep{AGENASPNE2024,epicentro_sdo}. 
While administrative data may be affected by heterogeneity in coding practices and periodic revisions to classification systems, they are generally considered reliable for identifying major surgical procedures. To minimise potential inconsistencies related to coding changes, the analysis is restricted to the period 2018–2024 \citep{MinSal_SDO}.

\subsection{Hip replacement admissions}
We focus on hip replacement procedures as a paradigmatic setting for the analysis of structured patient mobility. Hip arthroplasty is a high-volume, highly standardised intervention, predominantly performed on an elective or semi-elective basis \citep{ciminello2026total,berta2021hospitals}. In contrast to emergency care, the location of treatment is not dictated by urgency and proximity, allowing observed patient flows to reflect referral patterns, perceived differences in quality, and territorial attractiveness. We identify all ordinary inpatient admissions for hip replacement using ICD-9-CM procedure codes 81.51--81.53 and 00.70--00.87. The resulting dataset comprises 828,326 admissions over the period 2018--2024, involving residents from 109 ASLs across 20 Italian regions. Of these, 297,167 admissions (35.9$\%$ of total procedures) take place outside the patient’s ASL of residence. Annual mobility shares remain highly stable throughout the study period (Table~\ref{tab:admissions}), indicating that inter-ASL mobility in hip replacement constitutes a persistent structural feature of the healthcare system rather than a transient fluctuation.

\begin{table}[htbp]
\centering
\caption{Annual hip replacement admissions and inter-ASL mobility (2018--2024)}
\label{tab:admissions}
\begin{tabular}{@{}lrrrr@{}}
\toprule
Year & Total & Mobility & Mobility (\%) & Series (\%) \\
\midrule
2018 & 111,311 & 38,537 & 34.6 & 13.4 \\
2019 & 115,137 & 40,448 & 35.1 & 13.9 \\
2020 & 96,755 & 33,489 & 34.6 & 11.7 \\
2021 & 114,890 & 42,451 & 36.9 & 13.9 \\
2022 & 124,366 & 45,902 & 36.9 & 15.0 \\
2023 & 130,446 & 47,135 & 36.1 & 15.7 \\
2024 & 135,421 & 49,205 & 36.3 & 16.3 \\
\midrule
\textbf{Total} & \textbf{828,326} & \textbf{297,167} & \textbf{35.9} & \textbf{100.0} \\
\bottomrule
\end{tabular}
\end{table}

\subsection{Inter-area mobility patterns}

Macro-area aggregates reveal pronounced territorial asymmetries in patient mobility (Table~\ref{tab:macroareas}). Northern macro-areas exhibit limited inter-area mobility (3.4$\%$ in the North-West and 5.9$\%$ in the North-East), whereas residents in the South display substantially higher outward mobility, with 20.7$\%$ of admissions occurring outside their macro-area of residence. These patterns are consistent with, and further corroborate, the well-documented northward orientation of long-distance healthcare flows within Italy \citep{balia2018interregional,carnazza2025monetary}.
\begin{table}[htbp]
\centering
\caption{Inter-area mobility by macro-area (\% of admissions)}
\label{tab:macroareas}
\begin{tabular}{@{}lrr@{}}
\toprule
Macro-area & Intra-area & Inter-area \\
\midrule
North-West & 96.6 & 3.4 \\
North-East & 94.1 & 5.9 \\
Centre & 86.9 & 13.1 \\
South & 79.3 & 20.7 \\
Islands & 89.6 & 10.4 \\
\bottomrule
\end{tabular}
\end{table}

\begin{figure}[htbp]
\centering
\includegraphics[width=0.95\textwidth]{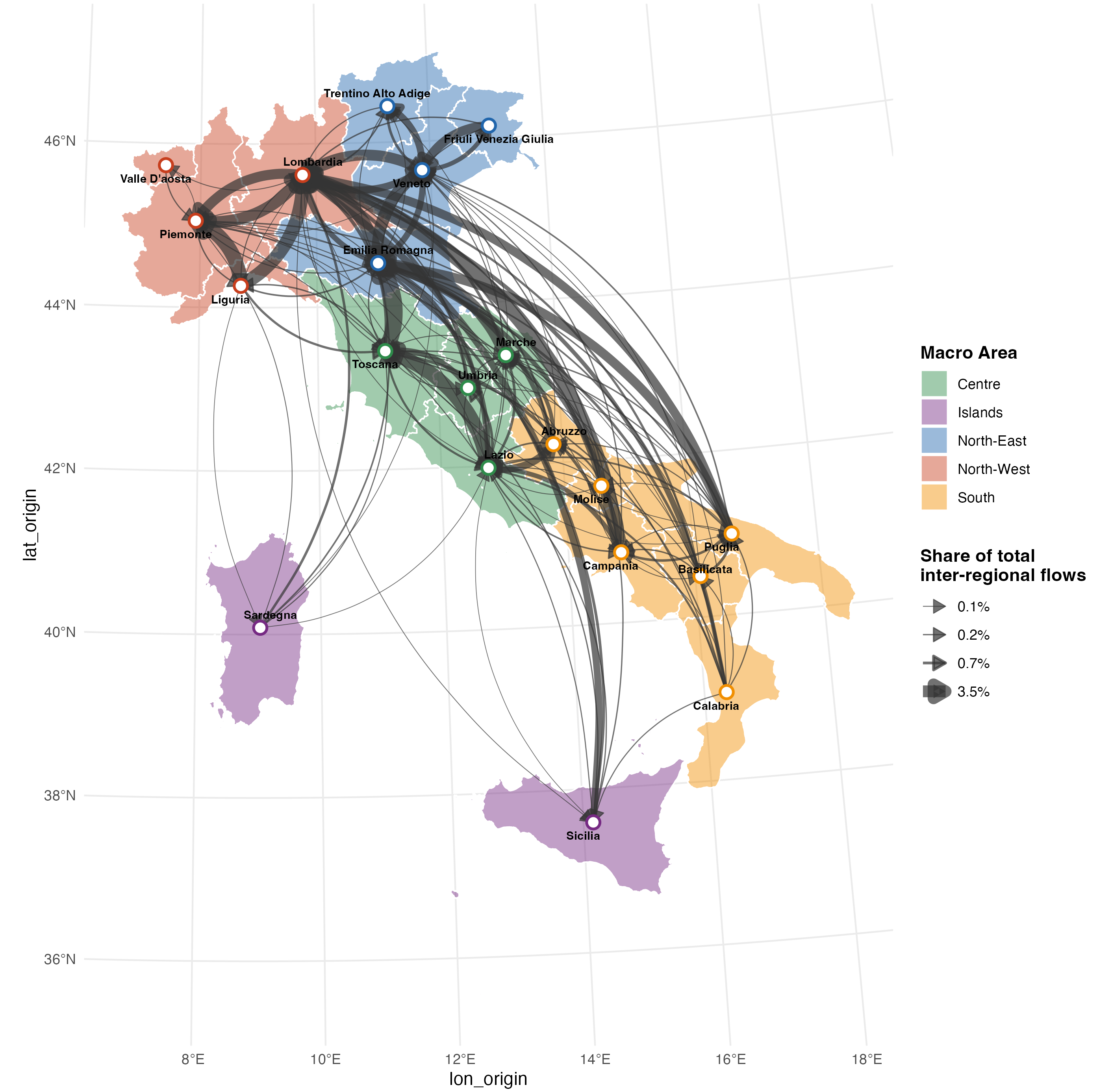}
\caption{Inter-regional hip replacement mobility flows (2018--2024). Arrows: residence $\rightarrow$ treatment. Width $\propto$ share of total inter-regional flows. Regions colored by macro-area.}
\label{fig:mobility_map}
\end{figure}

Figure~\ref{fig:mobility_map} illustrates the spatial distribution of outgoing mobility rates at the ASL level, highlighting marked heterogeneity across territories and a concentration of high-mobility areas in the South. The map also shows a clear directional pattern, with patient flows predominantly oriented towards northern regions—particularly Lombardia and Emilia-Romagna—which emerge as major destination hubs for patients from across the country.
Last, several southern regions are characterised by substantial outbound flows directed primarily towards central and northern Italy, while northern regions exhibit comparatively limited outward mobility and a stronger capacity to retain patients locally.

\subsection{ASLs road-network distances}
Geographical distance constitutes a key structural constraint on patient mobility. Even in the presence of quality differentials or established referral networks, travel entails economic, temporal, and organisational costs \citep{guarducci2022inter}. For elective procedures such as hip replacement, observed flows can therefore be interpreted as reflecting a trade-off between spatial proximity and provider attractiveness. In the absence of official inter-ASL distance measures, we compute road-network distances rather than Euclidean distances. Road distances provide a more realistic approximation of travel burden, as they account for transport network structure and routing constraints.
ASLs were geocoded using the Google Maps Geocoding API via the \texttt{googleway} package in \texttt{R} (where ASL barycentres are approximated by the nearest location returned by the Google API). For each ordered pair of distinct ASLs $(i,j)$, driving distances were obtained from the Google Directions API, specifying travel mode as \emph{driving} \citep{googleway,apparicio2008comparing,guarducci2022inter}. Total distance (in kilometres) was extracted for each route, with manual verification performed for a small number of ambiguous ASL identifiers. The resulting distances were then organised into a symmetric matrix. Road-network distances across observed mobility episodes exhibit marked right-skewness. The median travel distance is approximately 90 km, compared with a mean of about 190 km, indicating a substantial upper tail. Approximately 25$\%$ of mobility episodes occur within 50 km, while the maximum observed distance exceeds 1,700 km. These patterns point to the coexistence of short-range, proximity-driven exchanges and long-distance interregional mobility corridors.

\subsection{ASLs origin–destination matrices}
\label{sec:OD_matrices}
For each year $t=2018,\dots,2024$, we construct a directed origin--destination matrix
\[
\mathbf{Y}^{(t)} = (y_{ij}^{(t)}),
\]
where $y_{ij}^{(t)}$ denotes the number of residents in ASL $i$ treated in hospitals located in ASL $j$.
Diagonal elements ($i=j$) correspond to local admissions, while off-diagonal entries ($i \neq j$) capture inter-ASL mobility. As the analysis focuses on cross-territorial mobility, all subsequent analyses are restricted to off-diagonal entries.

The yearly origin–destination matrices are highly sparse. Table~\ref{tab:structure} summarises the distributional properties of the off-diagonal elements. Between 71\% and 79\% of all potential inter-ASL dyads exhibit zero flows in a given year, implying persistently low network density over the study period.
Conditional on being positive, flows are strongly right-skewed. Mean positive counts range from 2.87 to 4.19, while standard deviations are four to six times larger, and maximum values approach 700. The variance-to-mean ratio exceeds 100 in most years, indicating substantial overdispersion in the distribution of flows.

\begin{table}[htbp]
\centering
\caption{Structural features of yearly inter-ASL flow matrices (off-diagonal, $i \neq j$)}
\label{tab:structure}
\begin{tabular}{lrrrrrr}
\toprule
Year & Prop. zero & Mean & SD & Max count & Variance & Density \\
\midrule
2018 & 0.79 & 2.88 & 18.31 & 637 & 335.42 & 0.21 \\
2019 & 0.79 & 3.02 & 19.52 & 712 & 380.95 & 0.21 \\
2020 & 0.81 & 2.47 & 16.77 & 628 & 281.37 & 0.19 \\
2021 & 0.80 & 3.17 & 20.73 & 705 & 429.57 & 0.20 \\
2022 & 0.77 & 3.63 & 23.02 & 725 & 529.98 & 0.23 \\
2023 & 0.76 & 4.01 & 24.76 & 772 & 612.97 & 0.24 \\
2024 & 0.76 & 4.20 & 25.87 & 767 & 669.18 & 0.24 \\
\bottomrule
\end{tabular}
\end{table}

\section{The model}
\label{sec:model}
To capture the empirical features of the inter-ASL origin--destination matrices described in Section~\ref{sec:data}, we propose a dynamic latent space model for directed, weighted networks based on a hurdle negative binomial specification. The model decomposes the data-generating process into two components: a binary mechanism governing the presence of mobility flows, and a count component describing their magnitude. At the same time, each Local Health Authority (ASL) represents a node of the network, and it is embedded in a low-dimensional latent space that captures unobserved structural similarities between nodes.

Let $Y_{ij}^{(t)}$ denote the number of patients residing in ASL $i$ and treated in ASL $j$ during year $t$, for $i \neq j$, and let $\mathbf{Y}^{(t)} = (Y_{ij}^{(t)})$ denote the corresponding origin--destination matrix.

\subsection{Latent space representation and temporal dynamics}
Latent space models (LSMs) \citep{hoff2002latent} provide a probabilistic framework for relational data by embedding nodes in a low-dimensional Euclidean space. In this setting, each node is associated with an unobserved position, and the propensity to interaction between two nodes is modelled as a decreasing function of the Euclidean distance between their latent coordinates. This representation yields an interpretable geometric description of network structure and naturally captures features such as transitivity, clustering, and homophily \citep{hoff2002latent,handcock2007model}. Let $\mathbf{x}_i^{(t)} \in \mathbb{R}^d$ denote the latent position of ASL $i$ at time $t$. We fix $d=2$ to balance flexibility and interpretability, allowing direct visualisation of the latent configuration. The latent distance between ASLs $i$ and $j$ is defined as
\[
d_{ij}^{(t)} = \left\| \mathbf{x}_i^{(t)} - \mathbf{x}_j^{(t)} \right\|_2.
\]
Under this formulation, nearby ASLs in the latent space are expected to exchange larger volumes of patients, while distant pairs exhibit weaker interactions. The latent configuration therefore captures unobserved determinants of healthcare mobility, such as referral networks, perceived quality differences, institutional collaborations, or similarities in healthcare supply that are not directly observed in the data.
Healthcare mobility networks evolve over time due to changes in capacity, policy, and patient behaviour \citep{balia2018interregional}, making a static representation inadequate. To account for temporal dependence, we adopt the dynamic latent space model of \cite{sewell2015latent}. Let $\mathbf{X}^{(t)} = (\mathbf{x}_1^{(t)}, \dots, \mathbf{x}_N^{(t)})$ denote the matrix of latent positions at time $t$. For each ASL $i$,
\begin{equation}
\mathbf{x}_i^{(1)} \sim \mathcal{N}_2(\mathbf{0}, \sigma_X^2 I_2),
\qquad
\mathbf{x}_i^{(t)} \mid \mathbf{x}_i^{(t-1)} \sim \mathcal{N}_2\!\big(\mathbf{x}_i^{(t-1)}, \sigma_X^2 I_2\big),
\qquad t=2,\ldots,T,
    \label{eq:x}
\end{equation}

independently across nodes, where $\sigma_{X}^2$ controls the variability of latent trajectories. Compared with the more general specification in \cite{sewell2015latent}, we adopt a parsimonious formulation with isotropic covariance governed by a single scalar parameter $\sigma_{X}^2$, implying homogeneous variability across dimensions and reducing the number of parameters. This corresponds to a Gaussian random walk, where larger values of $\sigma_{X}^2$ allow substantial reconfiguration of the network structure between consecutive years, whereas smaller values enforce stronger temporal persistence. Under this formulation, the latent space defines a hidden dynamic process generating the observed sequence of mobility networks: conditional on the latent coordinates, networks are independent across time, and temporal dependence arises through the Markovian evolution of node positions.

\subsection{Modelling mobility flows}

We model patient flows using a hurdle specification \citep{mullahy1986specification}, which separates the mechanism governing the existence of a connection from that governing the magnitude of flows. The latter is modelled via a negative binomial, interpretable as a Poisson–Gamma mixture \citep{hilbe2011negative, cameron2013regression}, which accommodates the strong heterogeneity observed in origin--destination matrices, where a few dyads concentrate large volumes while many exhibit small counts. For each ordered pair $(i,j)$ and time $t$,
\begin{equation}
P(Y_{ij}^{(t)} = 0) = 1 - \pi_{ij}^{(t)}; \qquad 
Y_{ij}^{(t)} \mid Y_{ij}^{(t)} > 0 \sim \text{NegBin}\big(\mu_{ij}^{(t)}, r^{(t)}\big),
\label{eq:hurdle}
\end{equation}
where $\pi_{ij}^{(t)}$ is the probability of a positive flow, $\mu_{ij}^{(t)}$ is the conditional mean of the zero-truncated negative binomial, and $r^{(t)}$ is a dispersion parameter. Under this specification, the zero and positive counts are governed by separate processes: the binary component determines whether a flow is observed, while the truncated negative binomial models its magnitude given that it is positive. The probability of a positive flow is then specified via
\begin{equation}
    \text{logit}\big(\pi_{ij}^{(t)}\big) = \beta^{(t)},
    \label{eq:positiveflow}
\end{equation}
where $\beta^{(t)}$ is a  year-specific intercept capturing overall network density. This parsimonious formulation allows sparsity to vary over time while assuming homogeneous activation probabilities across dyads within each year. A negative binomial random 
variable with mean $\mu_{ij}^{(t)}$ and dispersion $r^{(t)}$ has variance 
$\mu_{ij}^{(t)} + {\mu_{ij}^{(t)2}}/{r^{(t)}}$, where smaller values of $r^{(t)}$ 
correspond to stronger overdispersion. Under the hurdle parameterisation in 
Equation~\ref{eq:hurdle}, the marginal mean and variance of $Y_{ij}^{(t)}$ are
\begin{equation}
    E[Y_{ij}^{(t)}] = \pi_{ij}^{(t)} \mu_{ij}^{(t)}, \qquad
    Var(Y_{ij}^{(t)}) = \pi_{ij}^{(t)} \mu_{ij}^{(t)} \left(1 + \mu_{ij}^{(t)} 
    \left(\frac{1}{r^{(t)}} + 1 - \pi_{ij}^{(t)}\right)\right).
\label{eq:mean_var}
\end{equation}
The mean is linked to the latent structure through:
\begin{equation}
\log \mu_{ij}^{(t)} = \alpha^{(t)} - d_{ij}^{(t)},
    \label{eq:mean}
\end{equation}
where $\alpha^{(t)}$ captures the overall level of mobility. This implies that expected flows decrease with latent distance, so that nearby ASLs exchange larger volumes of patients.

\subsection{Model extensions}
While the baseline formulation captures latent structural proximity, additional mechanisms may influence mobility flows. We therefore consider a set of extensions.

\subsubsection{Sender and receiver effects}
Healthcare mobility networks are inherently asymmetric: some territories tend to generate large numbers of outgoing patients, while others systematically attract them due to the presence of specialized hospitals or higher perceived quality of care.
To account for directional heterogeneity, we introduce node-specific sender and receiver effects \citep{holland1981exponential, van2004p2}, following the specification by \citet{d2020modeling}:
\begin{equation}
\log \mu_{ij}^{(t)} =
\alpha^{(t)} \cdot \frac{\gamma_i + \theta_j}{2} - d_{ij}^{(t)}.
\label{eq:sender_receiver}
\end{equation}
Here, $\gamma_i$ captures the propensity of ASL $i$ to generate outgoing patients, while $\theta_j$ measures the attractiveness of ASL $j$.

\subsubsection{Geographical distance}

Geographical distance may exert an additional deterrent effect beyond latent similarity, as travelling costs, accessibility constraints, and administrative borders often discourage long-distance treatment-seeking behaviour. We incorporate this in the hurdle component:
\begin{equation}
    \text{logit}\big(\pi_{ij}^{(t)}\big) = \beta^{(t)} - \rho^{(t)} G_{ij},
    \label{eq:geo_dist}
\end{equation}
where $G_{ij}$ denotes geographical distance between ASLs $i$ and $j$  as defined in Section~\ref{sec:data}, and $\rho^{(t)}$ measures the strength of the distance deterrence effect. 

\subsubsection{Exposure adjustment} 
Observed mobility flows are partly determined by the marginal volume of hospital activity within each territorial unit. ASLs with larger resident populations systematically generate a higher volume of outgoing patients, whereas areas that host major hospital facilities attract greater numbers of incoming cases. To account for such differences in scale across territories, we introduce an exposure term based on expected flows under independence, which is estimated from the data \citep{feng2022zero}:
\[
E_{ij}^{(t)} =
O_{i.}^{(t)} \frac{O_{.j}^{(t)}}{O^{(t)}}.
\]
The mean specification from Equation \ref{eq:mean} then becomes:
\begin{equation}
    \log \mu_{ij}^{(t)} =
\log E_{ij}^{(t)} + \alpha^{(t)} - d_{ij}^{(t)}.
\label{eq:offset}
\end{equation}
This extension allows the model to distinguish structural deviations from the baseline mobility patterns implied by marginal hospital activity from simple differences in territorial capacity and flow volume.

\subsubsection{Combined specifications}
\label{sec:specifications}
The above components define a family of nested models for the mean of positive flows. These models progressively incorporate additional structural mechanisms affecting healthcare mobility flows, and they are summarised in Table~\ref{tab:model_specs}. Each specification can additionally include geographical distance in the hurdle component, so that the probability of observing a positive flow is modelled as: $\text{logit}(\pi_{ij}^{(t)}) = \beta^{(t)} + \rho^{(t)} G_{ij}$.

\begin{table}[htbp]
\centering
\caption{Alternative specifications for the positive-flow mean}
\label{tab:model_specs}
\begin{tabular}{ll}
\toprule
Model & Mean specification \\
\midrule

1. Baseline &
$\log \mu_{ij}^{(t)} = \alpha^{(t)} - d_{ij}^{(t)}$ \\

2. Sender/receiver effects &
$\log \mu_{ij}^{(t)} =
\alpha^{(t)} \cdot \frac{\gamma_i + \theta_j}{2} - d_{ij}^{(t)}$ \\

3. Baseline + exposure adjustment &
$\log \mu_{ij}^{(t)} =
\log E_{ij}^{(t)} + \alpha^{(t)} - d_{ij}^{(t)}$ \\

4. Sender/receiver + exposure adjustment &
$\log \mu_{ij}^{(t)} =
\log E_{ij}^{(t)} + \alpha^{(t)} \cdot
\frac{\gamma_i + \theta_j}{2} - d_{ij}^{(t)}$ \\

\bottomrule
\end{tabular}
\end{table}

\subsection{Parameter estimation}
Estimation for the proposed modelling framework is carried out within a Bayesian framework. While the predictors entering the count and the hurdle components vary across specifications (see Section~\ref{sec:specifications}), all nested models share the same dynamic latent space structure, general prior specification scheme, and MCMC estimation strategy. 

\subsubsection{Likelihood and posterior distribution} 
Let $\Psi$ denote the collection of model parameters, including the latent coordinates and the scalar parameters governing both the hurdle and the count components. Let also $\text{logit}\big(\pi_{ij}^{(t)}\big)=\eta_{ij}^{(t)}$, $\log \mu_{ij}^{(t)}=\xi_{ij}^{(t)}$, where $\eta_{ij}^{(t)}$ and $\xi_{ij}^{(t)}$ are specified according to the chosen model formulation. Then, the likelihood can be written as:
\begin{equation}
    L(\Psi\mid \mathbf{Y})
=
\prod_{t=1}^{T}
\prod_{i\neq j}
\Bigg[
(1-\pi_{ij}^{(t)})^{\mathbbm{1}(Y_{ij}^{(t)}=0)}
\cdot
\left(
\pi_{ij}^{(t)}
\frac{f_{\mathrm{NB}}(Y_{ij}^{(t)};\mu_{ij}^{(t)},r^{(t)})}
{1-f_{\mathrm{NB}}(0;\mu_{ij}^{(t)},r^{(t)})}
\right)^{\mathbbm{1}(Y_{ij}^{(t)}>0)}
\Bigg].
\label{eq:likelihood}
\end{equation}
To account for temporal dependence, the latent coordinates are assigned a Gaussian random-walk prior (see Equation \ref{eq:x}). We assign conditionally independent Gaussian priors to the temporal parameters common to all specifications, \(\alpha^{(t)} \sim \mathcal{N}(\mu_\alpha,\sigma_\alpha^2)\) and \(\beta^{(t)} \sim \mathcal{N}(\mu_\beta,\sigma_\beta^2)\), \(t=1,\ldots,T\). When included, additional time-varying coefficient is assigned an independent Gaussian prior, namely \(\rho^{(t)} \sim \mathcal{N}(\mu_\rho,\sigma_\rho^2)\), \(t=1,\ldots,T\), while sender and receiver effects satisfy \(\gamma_i \sim \mathcal{N}(0,\sigma_\gamma^2)\) and \(\theta_j \sim \mathcal{N}(0,\sigma_\theta^2)\), \(i,j=1,\ldots,n\). To ensure positivity of the negative binomial dispersion parameter, we use the reparameterization \(a^{(t)}=1/\sqrt{r^{(t)}}\), \(t=1,\ldots,T\), and assume \(a^{(t)} \sim \mathcal{N}^{+}(0,\sigma_a^2)\), which induces a proper prior on \(r^{(t)}\) over the positive real line.

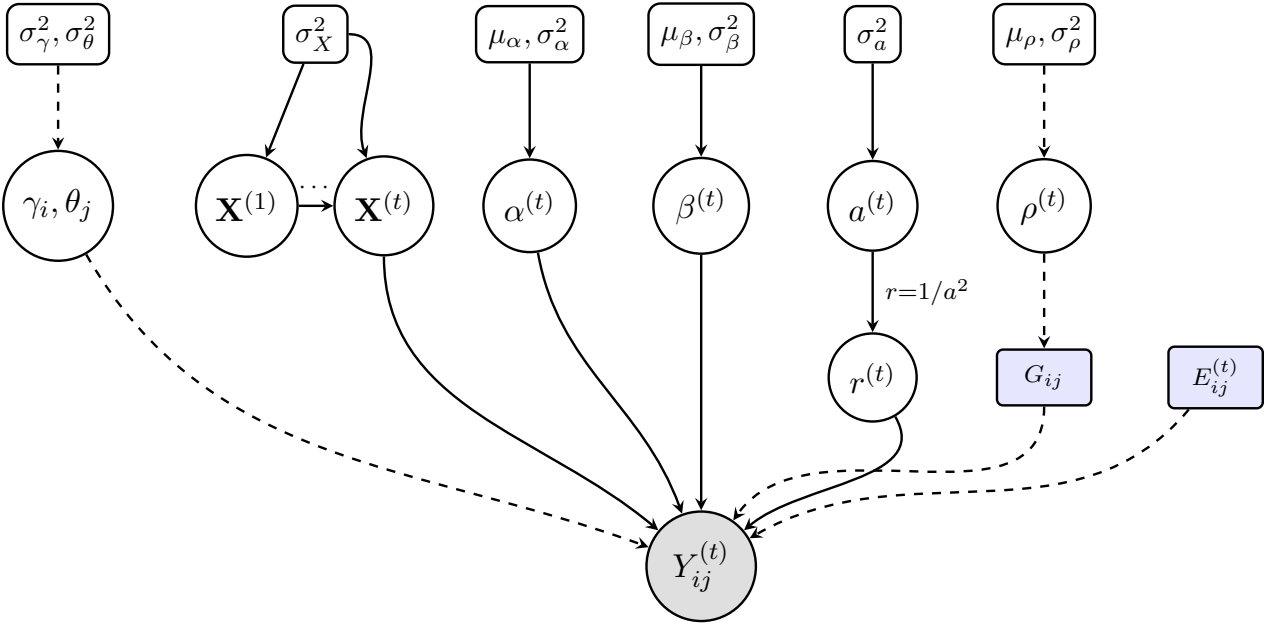
\begin{figure}[htbp]
\centering
\resizebox{\textwidth}{!}{
\begin{tikzpicture}[
    latent/.style={circle, draw, thick, minimum size=0.9cm, font=\small},
    obs/.style={circle, draw, thick, fill=gray!25, minimum size=0.9cm, font=\small},
    data/.style={rectangle, draw, thick, fill=blue!10, rounded corners=2pt, minimum width=1.1cm, minimum height=0.65cm, font=\scriptsize},
    hyper/.style={rectangle, draw, thick, rounded corners,
                  inner sep=4pt, minimum height=0.65cm, font=\footnotesize},
    arrow/.style={->, >=stealth, thick},
    optarrow/.style={->, >=stealth, thick, dashed, black}
]

\node[hyper] (hst) at (-1.0,  0) {$\sigma_\gamma^2,\sigma_\theta^2$};
\node[hyper] (sX)  at ( 2.0,  0) {$\sigma_X^2$};
\node[hyper] (ha)  at ( 4.5,  0) {$\mu_\alpha,\sigma_\alpha^2$};
\node[hyper] (hb)  at ( 6.5,  0) {$\mu_\beta,\sigma_\beta^2$};
\node[hyper] (hsa) at ( 8.5,  0) {$\sigma_a^2$};
\node[hyper] (hr)  at (10.5,  0) {$\mu_\rho,\sigma_\rho^2$};
\node[latent] (sr) at (-1.0, -2) {$\gamma_i,\theta_j$};
\node[latent] (x1) at ( 1.2, -2) {$\mathbf{X}^{(1)}$};
\node[latent] (xt) at ( 2.8, -2) {$\mathbf{X}^{(t)}$};
\node[latent] (al) at ( 4.5, -2) {$\alpha^{(t)}$};
\node[latent] (be) at ( 6.5, -2) {$\beta^{(t)}$};
\node[latent] (a)  at ( 8.5, -2) {$a^{(t)}$};
\node[latent] (rh) at (10.5, -2) {$\rho^{(t)}$};
\node[latent] (r)  at ( 8.5, -4) {$r^{(t)}$};
\node[data]   (G)  at (10.5, -4) {$G_{ij}$};
\node[data]   (E)  at (12.5, -4) {$E_{ij}^{(t)}$};
\node[obs] (Y) at (6.5, -6.2) {$Y_{ij}^{(t)}$};
\draw[arrow] (sX) -- (x1);
\draw[arrow] (sX) to[out=0, in=110] (xt);
\draw[arrow] (x1) -- node[above, font=\scriptsize] {$\cdots$} (xt);
\draw[arrow] (ha)  -- (al);
\draw[arrow] (hb)  -- (be);
\draw[arrow] (hsa) -- (a);
\draw[arrow] (a) -- node[right, font=\scriptsize] {$r{=}1/a^2$} (r);
\draw[arrow] (xt) to[out=-90, in=140] (Y);
\draw[arrow] (al) to[out=-80, in=110] (Y);
\draw[arrow] (be) -- (Y);
\draw[arrow] (r)  to[out=-60, in=40]  (Y);
\draw[optarrow] (hst) -- (sr);
\draw[optarrow] (sr)  to[out=-60, in=160] (Y);
\draw[optarrow] (hr) -- (rh);
\draw[optarrow] (rh) -- (G);
\draw[optarrow] (G)  to[out=-90, in=55] (Y);
\draw[optarrow] (E)  to[out=-130, in=30] (Y);
\end{tikzpicture}
}
\caption{Hierarchy structure of the model.}
\label{fig:dag}
\end{figure}

Under this specification, the posterior distribution of the unknown quantities is, up to a normalizing constant,
\begin{equation}
p(\Psi \mid \mathbf{Y})
\propto
L(\Psi \mid \mathbf{Y})
\,
f_{\mathbf{X}}(\mathbf{X})
\,
f_{\boldsymbol{\alpha}}(\boldsymbol{\alpha})
\,
f_{\boldsymbol{\beta}}(\boldsymbol{\beta})
\,
f_{\boldsymbol{r}}(\boldsymbol{r})
\,
f_{\boldsymbol{\omega}}(\boldsymbol{\omega}),
\label{eq:posterior}
\end{equation}

where $L(\Psi \mid \mathbf{Y})$ is the hurdle negative binomial likelihood defined above, $f_{\mathbf{X}}(\mathbf{X})$ denotes the dynamic Gaussian prior for the latent coordinates, $f_{\boldsymbol{\alpha}}(\boldsymbol{\alpha})$ and $f_{\boldsymbol{\beta}}(\boldsymbol{\beta})$ denote the Gaussian priors for the common time--varying parameters, $f_{\boldsymbol{r}}(\boldsymbol{r})$ is the prior induced by the half-normal specification on $a^{(t)}$, and $f_{\boldsymbol{\omega}}(\boldsymbol{\omega})$ collects the priors of the additional parameters included in the selected model specification.

The resulting full conditional distributions are not generally available in closed form. For this reason, posterior inference is carried out by a Metropolis--Hastings within Gibbs algorithm, described in the following subsection.

\subsubsection{Model estimation} 
\label{sec:model_estimation}
At each iteration of the chain, the time-specific scalar parameters in the hurdle and the count components are updated using random-walk Metropolis–Hastings steps. In the baseline specification, this applies to the intercept and hurdle coefficient, while in more general specifications the same scheme extends to additional parameters, including sender and  receiver effects, and geographical effects.

Latent coordinates are updated separately, year by year and node by node, rather than jointly as a single block. This strategy allows the algorithm to adjust only poorly located positions while preserving coordinates that have already stabilized, providing greater flexibility and stability in practice.
Because the likelihood depends on pairwise distances between latent coordinates, it is invariant to translations and rotations of the latent configuration. To address this lack of identifiability, after each sweep of latent-coordinate updates the configuration is re-centered and aligned via a Procrustes transformation, ensuring comparability of latent trajectories across iterations and over time.

To guarantee positivity of the negative binomial dispersion parameter, the sampler operates on the reparameterization  \(a^{(t)} = 1/\sqrt{r^{(t)}}\), with Metropolis–Hastings updates performed on this transformed scale. Proposal variances are adaptively tuned during burn-in based on empirical acceptance rates, with separate adaptation for each parameter block.

Starting values are set using a data-driven initialization strategy. Initial latent coordinates are obtained from multidimensional scaling applied to a dissimilarity matrix derived from observed flows; hurdle coefficients are initialized from the empirical density of positive entries; and count-component parameters are initialized via a negative binomial regression based on the initial latent distances. Geographical and exposure parameters, when included, are set to neutral values, whereas sender and receiver effects are initialized from empirical summaries of outgoing and incoming flows.

When sender and receiver effects are included, identifiability is enforced by fixing one sender effect and one receiver effect to reference values throughout estimation. Posterior summaries are computed from the retained draws after discarding burn-in iterations.

\section{Simulation study}
\label{sec:simulation}
We conducted a simulation study to assess the ability of the proposed MCMC algorithm, both for the baseline model and the model including sender and receiver effects, to recover true parameter values under controlled conditions.
The study was specifically designed to evaluate the performance of the estimation procedure across varying network sizes, panel lengths, levels of sparsity, and degrees of overdispersion, reflecting the structural characteristics of the hospital mobility data analysed in Section~\ref{sec:data}.

\subsection{Simulation design}
Synthetic directed networks were generated following the data generating processes implied by each model specification (Section \ref{sec:model}). In all scenarios, latent positions at time $t=1$  were sampled from a standard bivariate normal distribution and evolved over time according to a Gaussian random walk with variance $\sigma^2_X = 0.1$. The intercepts $\alpha^{(t)}$ were set to increase linearly from 3.5 to 4.5 over the $T$ periods, reflecting the gradual growth pattern observed in the AGENAS data. For the multiplicative model, sender effects $\gamma_i$ and receiver effects $\theta_j$ were simulated from independent standard normal distributions. 

A factorial design was considered by crossing the following factors:
\begin{enumerate}
    \item Network size: $n \in \{100, 150\}$ nodes;
    
    \item Panel length: $T \in \{5, 10, 15\}$ time periods;
    
    \item Sparsity level: the hurdle intercept $\beta$ was calibrated to yield approximately 10\%, 30\%, and 50\% structural zeros;
    
    \item Overdispersion: the negative binomial dispersion parameter was set to $r = 5$ (low overdispersion) and $r = 0.5$ (high overdispersion).
\end{enumerate}

This yields $36$ simulation scenarios. For each configuration, $S=100$ independent datasets were generated. The two models were estimated using $30{,}000$ MCMC iterations, discarding the first $5{,}000$ as burn-in.

\subsection{Simulation results}
\label{sec:simulation_results}
We discuss the results by first examining the recovery of the latent positions, which represent the main inferential target of the model, and then assessing the estimation of the scalar parameters under each model specification.

\begin{figure}[htbp]
\centering
\includegraphics[width =1\textwidth]{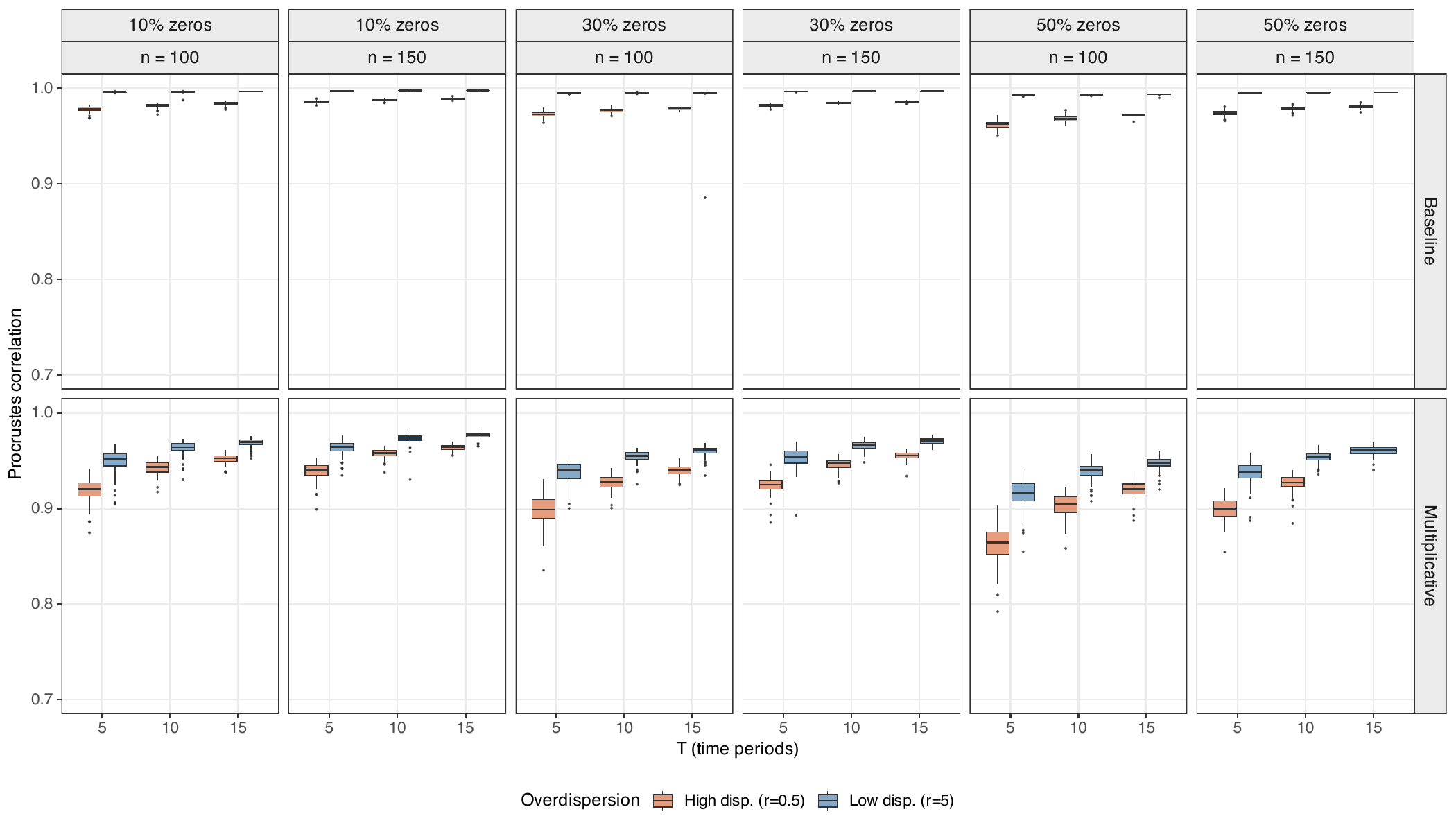}
\caption{Simulation study. Procrustes correlation between estimated posterior mean latent positions and true simulated positions, across all combinations of network size $n$, panel length $T$, sparsity level, and overdispersion, for the baseline model (top) and the multiplicative model (bottom).}
\label{fig:procrustes_simulation}
\end{figure}

Figure~\ref{fig:procrustes_simulation} reports the values of the Procrustes correlation between estimated (posterior mean) and simulated latent coordinates across all scenarios. 
Under the baseline specification, the recovery of the latent geometry is uniformly highly accurate, with correlations ranging from 0.96 to 1.00. 
Estimation accuracy improves with increasing network size and longer panel length, reflecting the greater amount of information available for identifying the underlying latent structure. 
Under the multiplicative specification (Model~2, Table~\ref{tab:model_specs}), the latent space is also recovered with high accuracy, although the correlations are slightly lower. 
However, recovery remains robust at all levels of sparsity and overdispersion.

\begin{figure}[htbp]
\centering
\includegraphics[width=1\textwidth]{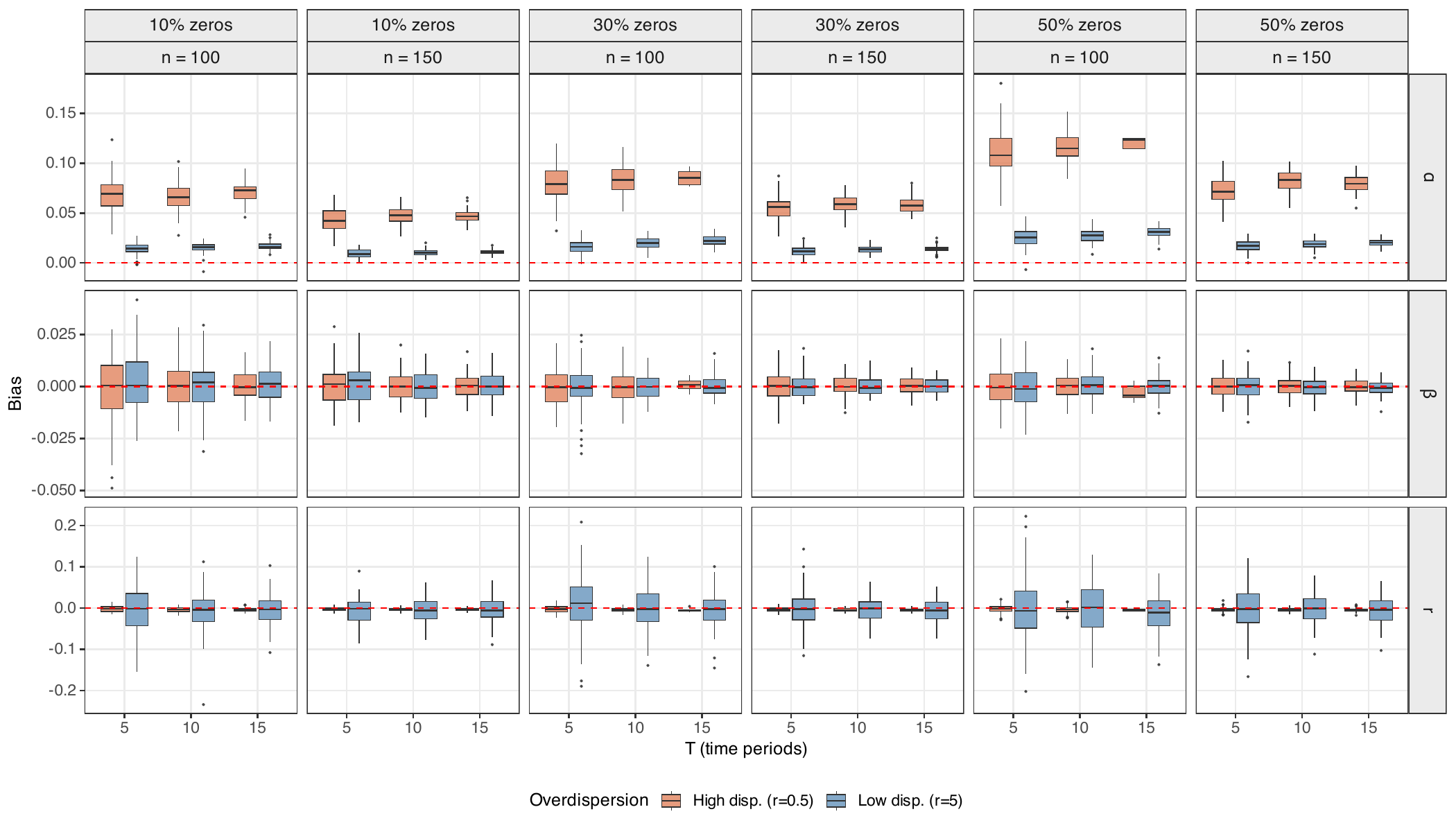}
\caption{Simulation study. Bias of posterior means for $\alpha^{(t)}$ (top), $\beta^{(t)}$ (middle), and $r^{(t)}$ (bottom) under the baseline model, across all combinations of network size $n$, panel length $T$, sparsity level, and overdispersion regime.}
\label{fig:bias_base_simulation}
\end{figure}

Figure~\ref{fig:bias_base_simulation} presents the posterior mean bias for $\alpha^{(t)}$, $\beta^{(t)}$, and $r^{(t)}$ in all scenarios under the baseline model. 
The hurdle intercept $\beta^{(t)}$ is recovered with negligible bias in every configuration, indicating that the zero-generating component is well identified. 
The dispersion parameter $r^{(t)}$ is also estimated accurately on average, with posterior uncertainty naturally larger under low overdispersion ($r=5$), when the negative binomial distribution approaches the Poisson limit.
The empirical coverage of the 95\% credible intervals for $\beta^{(t)}$ and $r^{(t)}$ is close to the nominal level in every setting examined (see Appendix).
The intercept $\alpha^{(t)}$ displays a systematic positive bias that increases with both sparsity and overdispersion. Under low overdispersion, the bias remains small (0.01--0.03).
Under high overdispersion ($r=0.5$), it becomes more 
pronounced, ranging from 0.04 to 0.12, and the empirical coverage of the 95\% credible interval falls substantially below the nominal level; full results are reported in the
Appendix. This systematic bias arises from the dependence of the objective function on purely latent quantities, which makes the magnitude of the intercept difficult to identify independently \citep{d2023model}.

\begin{figure}[htbp]
\centering
\includegraphics[width=1\textwidth]{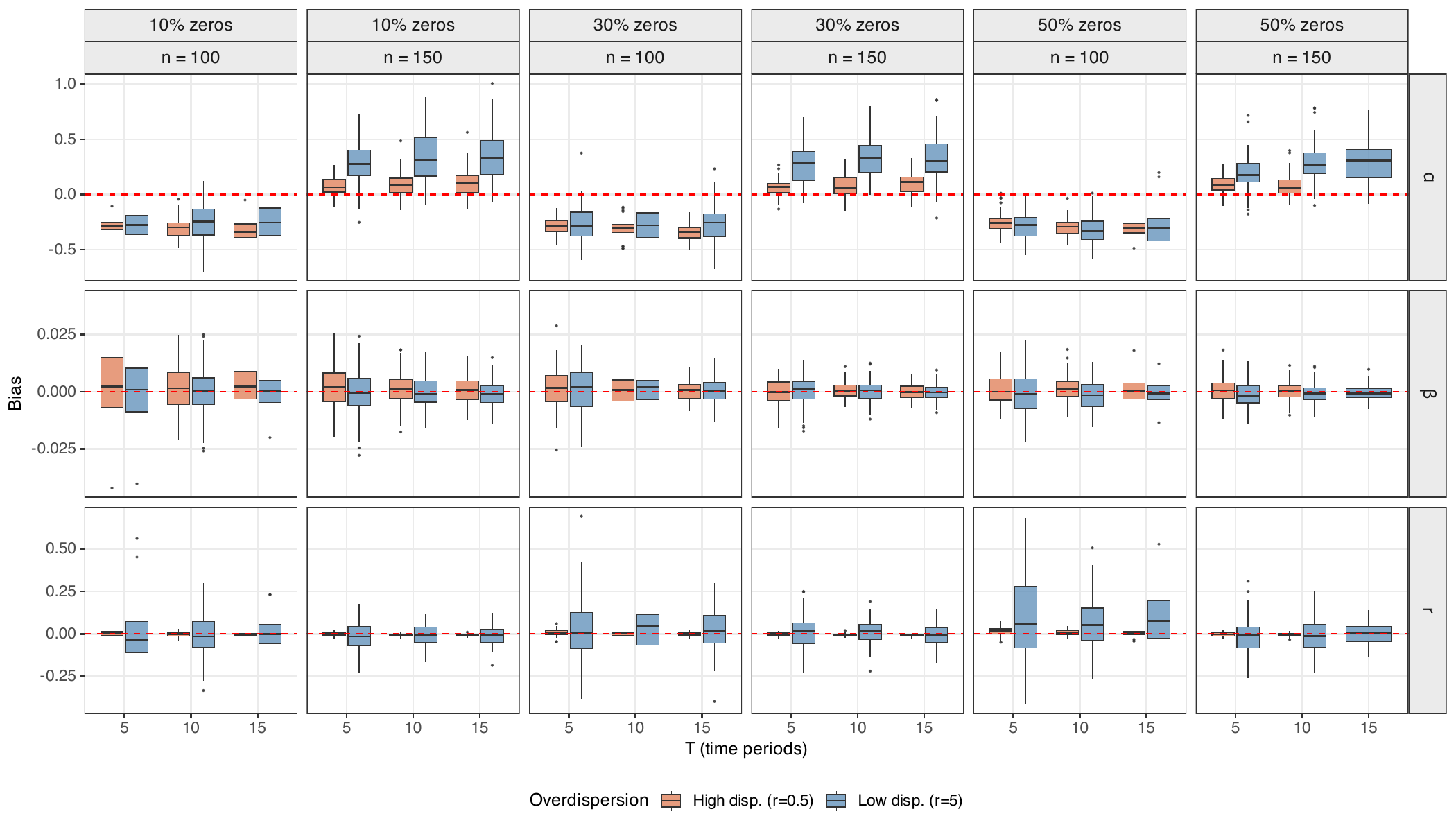}
\caption{Simulation study. Bias of posterior means for $\alpha^{(t)}$ (top), $\beta^{(t)}$ (middle), and $r^{(t)}$ (bottom) under the multiplicative model, across all combinations of network size $n$, panel length $T$, sparsity level, and overdispersion regime.}
\label{fig:bias_molt_simulation}
\end{figure}

Figure~\ref{fig:bias_molt_simulation} presents the corresponding bias results for the multiplicative specification.
The hurdle intercept $\beta^{(t)}$ and the dispersion parameter $r^{(t)}$ are again estimated with negligible bias and well-calibrated credible intervals across all scenarios, confirming that these components of the model are not affected by the additional complexity introduced by the sender and receiver effects.
The intercept $\alpha^{(t)}$, however, exhibits a systematic bias whose sign and magnitude depend on network size. 
This behaviour reflects a partial non-identifiability intrinsic to the multiplicative parametrisation, which is not fully resolved by the reference-node constraints imposed during estimation.

\begin{figure}[htbp]
\centering
\includegraphics[width=1\textwidth]{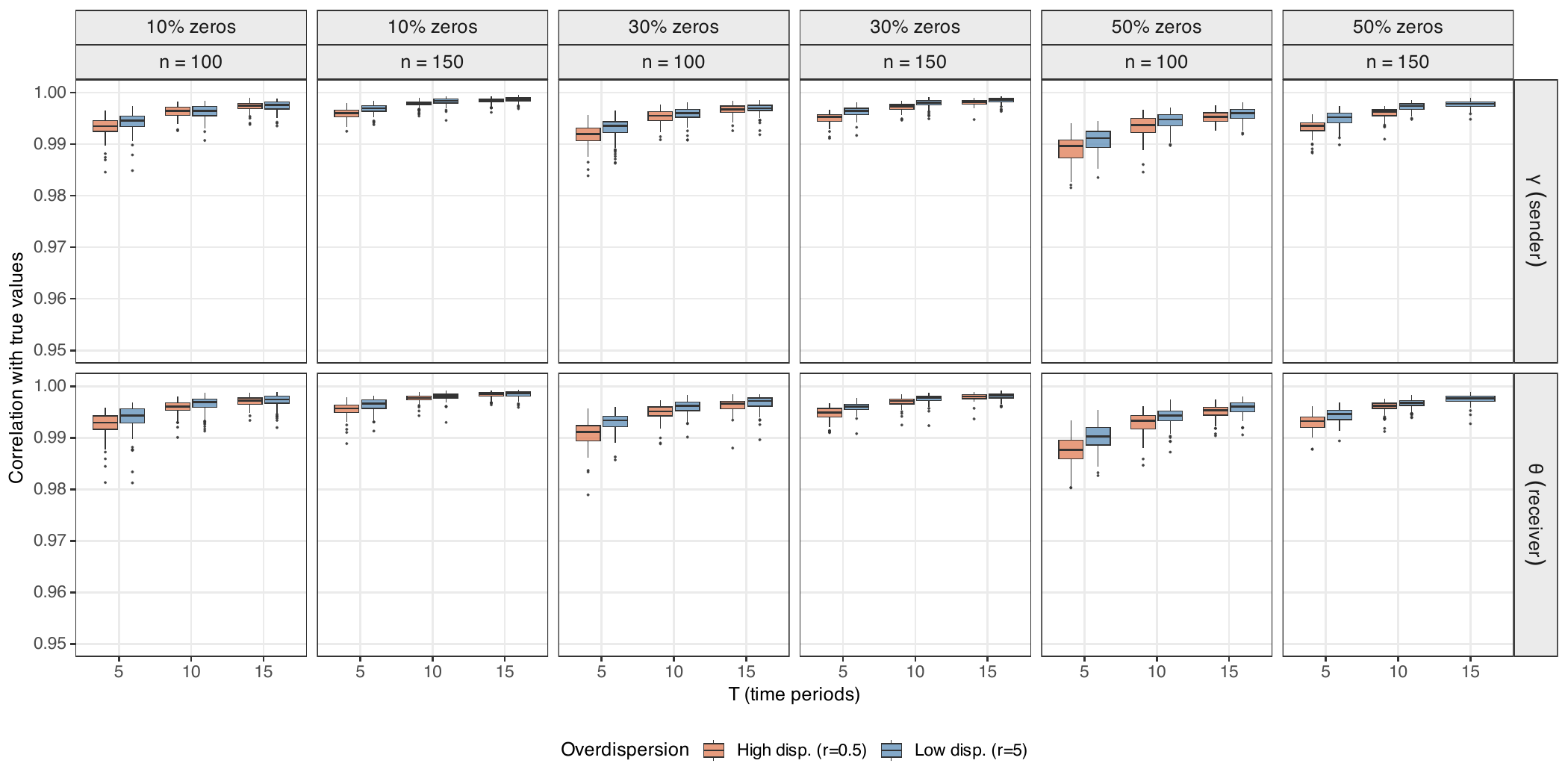}
\caption{Simulation study. Correlation between posterior means and true values for sender effects $\gamma_i$ (top) and receiver effects $\theta_j$ (bottom) under the multiplicative model, across all simulation scenarios.}
\label{fig:gamma_theta_simulation}
\end{figure}

Figure~\ref{fig:gamma_theta_simulation} reports the correlation between posterior means and true values for the sender effects $\gamma_i$ and receiver effects $\theta_j$. 
In all scenarios, the correlations are remarkably high, always exceeding 0.98 and often above 0.99. This indicates that the model successfully captures node-specific heterogeneity in sending and receiving behaviour even under high sparsity and strong overdispersion.

Overall, the results are satisfactory, and the simulation study demonstrates that the proposed methodology performs robustly across a broad range of data-generating scenarios. 
The latent space is consistently recovered with high accuracy, the hurdle and dispersion parameters are estimated with substantial reliability, and the multiplicative specification reproduces sender and receiver heterogeneity with fidelity. 
The primary limitations concern the estimation of the intercept parameter under high overdispersion in the baseline model and the scale non-identifiability affecting the intercept in the multiplicative specification. 
Nonetheless, these issues do not compromise the recovery of the latent network structure, which constitutes the primary inferential objective of the analysis.

\section{Application to mobility for hip replacement}
\label{sec:application}

We applied all model specifications described in Section~\ref{sec:model} to the inter-ASL hip replacement origin--destination matrices for the period 2018--2024 (see Section~\ref{sec:OD_matrices}). Each model was estimated via the MCMC algorithm described in Section~\ref{sec:model_estimation}, running $30{,}000$ iterations and discarding the first $5{,}000$ as burn-in. We start by comparing the eight specifications and selecting the preferred model, that is the one achieving the best fit according to the majority of selection criteria. We then analyse the estimated latent space, the sender and receiver effects, and the temporal dynamics of overdispersion.

\subsection{Model comparison and selection}
Table~\ref{tab:model_comparison_data} summarises the performance of all eight specifications across a battery of metrics: the Deviance Information Criterion (DIC), the mean absolute error (MAE) and the root mean squared error (RMSE) computed over all dyads and over positive flows only, cosine similarity between observed and replicated flow 
vectors, zero-flow accuracy, false positive and false negative rates.

\begin{table}[htbp]
\centering
\caption{Model comparison metrics for all eight specifications. 
For each column, the best value is highlighted in bold. 
The overall preferred model is additionally highlighted in bold in the model name column.}
\label{tab:model_comparison_data}
\resizebox{\textwidth}{!}{
\begin{tabular}{lrrrrrrrrr}
\toprule
Model & DIC & MAE & RMSE & MAE$^+$ & RMSE$^+$ & Cosine & ZeroAcc & FPR & FNR \\
\midrule
Base               & 186842.6 & 5.29 & 28.31 & 16.60 & 50.18 & 0.07 & 65.82 & 21.93 & 78.07 \\
Base\_Geo          & 180290.1 & 5.51 & 29.59 & 17.36 & 52.72 & 0.11 & 68.96 & 19.91 & \textbf{70.92} \\
Base\_Offset       & 181672.7 & 3.99 & \textbf{22.81} & 15.58 & \textbf{47.63} & 0.19 & 65.81 & 21.64 & 78.07 \\
Base\_Geo\_Offset  & 175344.8 & 3.97 & 23.23 & 15.56 & 48.46 & 0.28 & 68.94 & 19.92 & 70.94 \\
Mult               & 180224.1 & 4.06 & 25.61 & 15.58 & 51.44 & 0.21 & 65.81 & 21.94 & 78.07 \\
Mult\_Geo          & 173790.1 & 4.05 & 27.06 & 15.71 & 55.45 & 0.30 & 68.94 & 19.93 & 70.92 \\
Mult\_Offset       & 180046.5 & 3.96 & 24.12 & \textbf{15.49} & 49.68 & 0.21 & 65.80 & 21.95 & 78.11 \\
\textbf{Mult\_Geo\_Offset}  & \textbf{173727.5} & \textbf{3.95} & 25.96 & 15.60 & 53.53 & \textbf{0.32} & \textbf{68.94} & \textbf{19.91} & 70.95 \\
\bottomrule
\end{tabular}}
\end{table}

The multiplicative sender--receiver model with geographic distance in the hurdle component and exposure adjustment (\texttt{Mult\_Geo\_Offset}) provides the best overall fit. 
It achieves the lowest DIC among all specifications, indicating a good trade-off between fit and complexity. It also attains the highest cosine similarity, reflecting the most accurate reproduction of the relational structure of the mobility network. 
The false positive rate ($19.91\%$) and zero-flow accuracy ($68.94\%$) are also competitive.

The baseline model with exposure adjustment and geographical distance (\texttt{Base\_Geo\_Offset}) emerges as a strong competitor in terms of point prediction, yielding lower values for RMSE, MAE$^+$, and RMSE$^+$ across all dyads. 
However, it produces a higher DIC ($175{,}344.8$) and lower cosine similarity ($0.28$), indicating that it optimises the prediction of individual flow counts at the expense of capturing the overall network structure. Since the primary goal of this analysis is to describe the organisation of the mobility network rather than to minimise prediction error for individual dyads, we select \texttt{Mult\_Geo\_Offset} as the preferred specification. 
Nonetheless, \texttt{Base\_Geo\_Offset} constitutes a relevant alternative whose results are reported in full in the Appendix.

An elevated false negative rate emerges across all specifications. This is not entirely surprising in a setting characterised by pronounced sparsity and overdispersion, where many flows are very small and can be misclassified as zeros. In particular, dyads with only one or two patients may occasionally be estimated as zero by the model, thus inflating the FNR without necessarily compromising the overall interpretation of the network. The main objective of this application is to recover the latent structure of inter-ASL mobility and the aggregate pattern of flows, rather than to achieve exact prediction of small counts. 
Consequently, the MAE remains limited compared with the empirical variability of the data, suggesting that the model captures the main mobility patterns adequately.

To further reduce the false negative rate, we also considered replacing the logistic hurdle component with a generalised additive model \citep{hastie2017generalized}, which would allow the probability of observing a positive flow to vary non-linearly with geographical distance.
Implemented as a two-stage plug-in procedure \citep{vermunt2010latent} -- in which the probabilities estimated by the GAM are treated as fixed inputs to the count component -- this approach would also simplify the MCMC sampler, since the hurdle parameters would no longer need to be updated iteratively. 
The GAM-based models produce the lowest DIC values overall, ranging from 165,782.1 (\texttt{Mult\_Geo\_GAM}) to 172,743.3 (\texttt{Base\_Geo\_GAM}), and lower the false negative rate to roughly 62\%, compared with 71\% under the preferred logistic formulation.
However, this gain in fit does not materially affect the latent network structure or the sender–receiver effects: Procrustes correlations between posterior mean configurations are above 0.998 in every year 
of the study, and the Pearson correlations for sender and receiver effects are 0.9996 and 0.9999, respectively.
Thus, the choice of hurdle component does not change the substantive findings of the analysis.
We nevertheless adopt the standard logistic specification for all subsequent results, because it provides a fully 
parametric and interpretable model whose parameters 
can be directly subjected to posterior inference. 
We leave the GAM extension as an avenue for future research.

\subsection{Latent geometry of inter-ASL mobility}

\begin{figure}[htbp]
\centering
\includegraphics[width=0.95\textwidth]{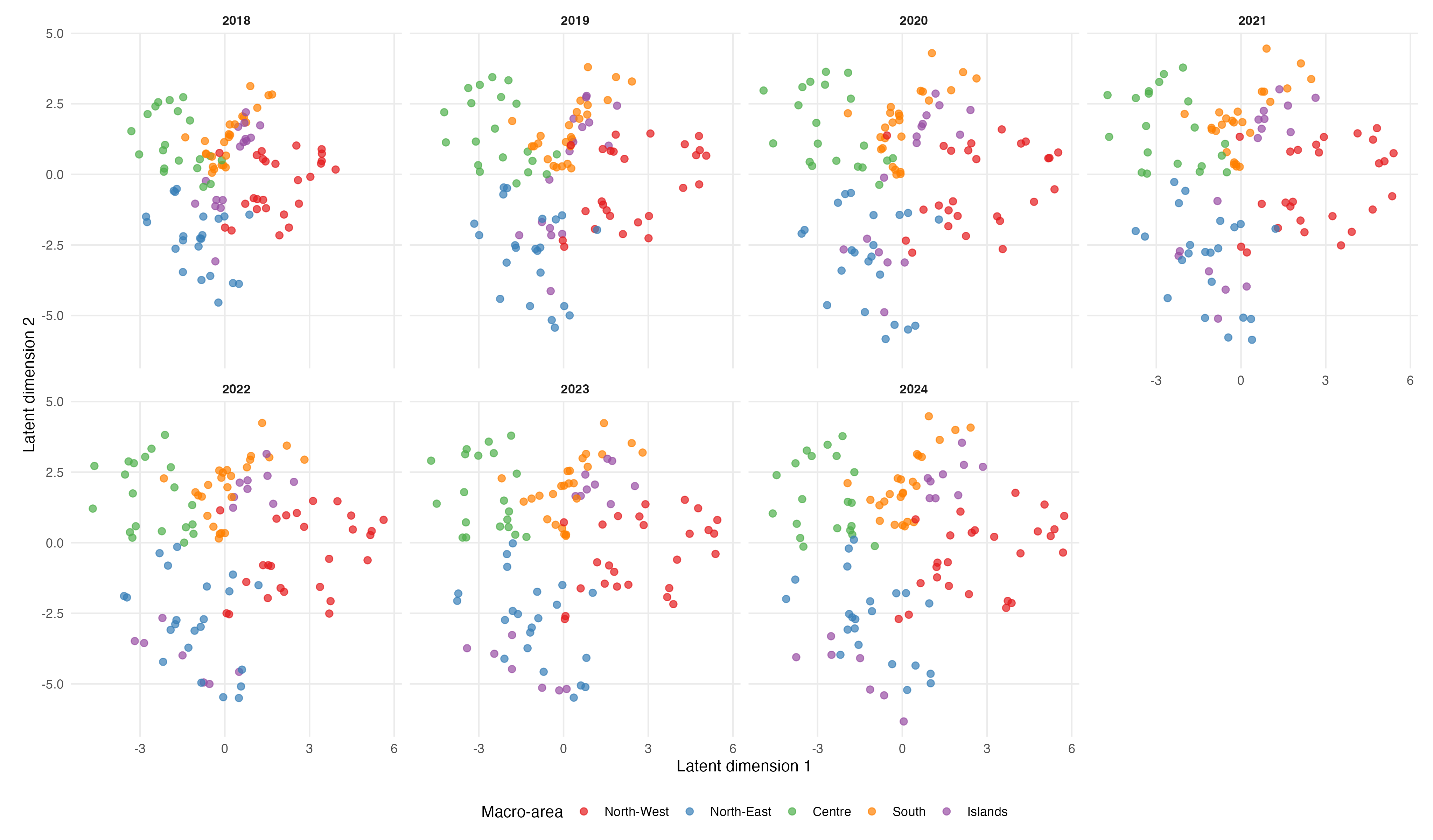}
\caption{Temporal evolution of latent positions for 109 Italian ASLs (2018–2024) coloured by macro-area.}
\label{fig:latent_macroarea}
\end{figure}

Figure~\ref{fig:latent_macroarea} shows the estimated posterior mean latent positions for all 109 ASLs across the seven years of the study period, coloured by macro-area.
A clear and persistent clustering by broad territorial area is visible throughout the observation period: ASLs from the same macro-area tend to occupy nearby positions in the latent space, consistent with the predominance of within-area mobility over long-distance inter-area flows. The latent representation therefore captures a meaningful aspect of the structure of the Italian healthcare mobility network, combining geographic regularities with unobserved institutional and referral similarities.

The two island regions — Sicilia and Sardegna — form two clearly distinct clusters that are systematically separated from one another. This suggests that residents of the islands do not exhibit a common pattern of healthcare-seeking behaviour; instead, each island is oriented toward a different group of mainland ASLs, probably reflecting the different logistical and infrastructural conditions.

\begin{figure}[htbp]
\centering
\includegraphics[width=1\textwidth]{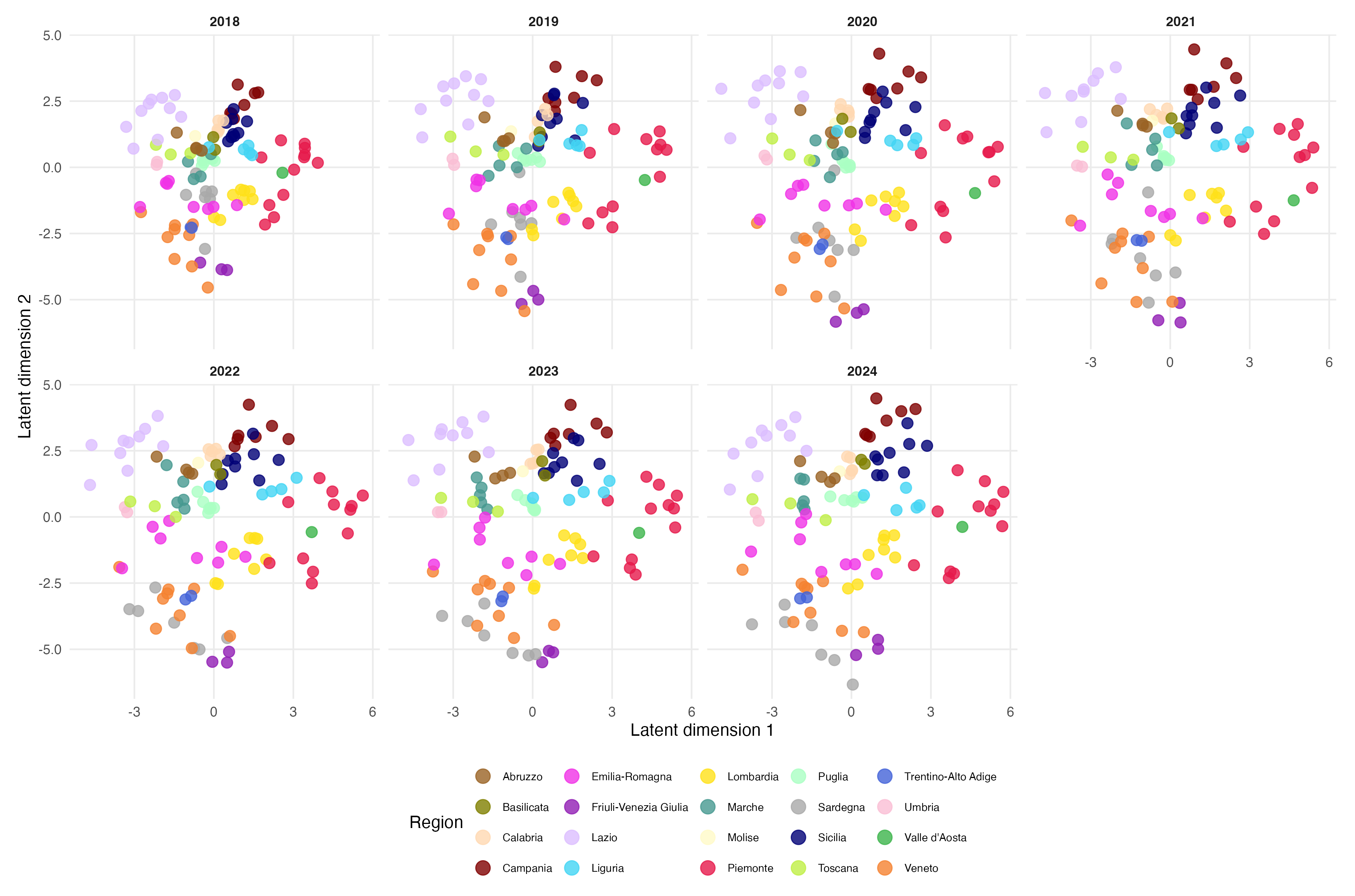}
\caption{Temporal evolution of latent positions for 109 Italian ASLs (2018–2024) coloured by region.}
\label{fig:latent_region}
\end{figure}

When nodes are coloured by region (Figure~\ref{fig:latent_region}), further sub-regional patterns emerge. Sardinian ASLs consistently lie closer to north-eastern regions such as Veneto and Friuli-Venezia Giulia, possibly reflecting referral ties and transport connections that make those facilities more accessible than many geographically-closer mainland destinations. Sicilian ASLs, by contrast, show greater latent proximity 
to the north-western region of Liguria and to southern mainland areas such as Campania, Calabria, and Basilicata. 

The dispersion of latent positions---measured as the mean Euclidean distance from the centroid---increases monotonically across the study period, from 2.06 in 2018 to 3.16 in 2024. The comparatively low dispersion observed in 2018 may partly reflect residual data quality issues associated with the transition from 2017, a year that was excluded from the analysis due to known inconsistencies in the discharge records; if these issues were not fully resolved in the first year of the sample, the 2018 configuration would be estimated with somewhat less precision, yielding a more compressed latent geometry. 
The acceleration in dispersion between 2019 and 2021---from 2.63 to 3.02---coincides with the disruption to elective surgical activity induced by the COVID-19 pandemic.
Elective orthopaedic surgery, including hip replacement, was suspended or severely reduced across Italian hospitals during the national lockdown of spring 2020 \citep{placella2020covid, d2020disruption}. The subsequent recovery was spatially unequal, with volumes returning to pre-pandemic levels faster in some regions than others \citep{di2024equity}, amplifying latent differentiation among ASLs.

The continued increase in dispersion through 2022--2024 suggests that this differentiation was not transient but reflects a structural reorganisation of mobility patterns in the post-pandemic period. A plausible contributing factor is the differential implementation of the National Recovery and Resilience Plan (\textit{Piano Nazionale di Ripresa e Resilienza} -- PNRR), which allocated \euro{}15.63 billion to the Italian NHS under Mission 6, including substantial investments in hospital infrastructure and technological modernisation \citep{cascini2023national, brambilla2022new}. 
To the extent that these investments were absorbed unevenly across regions, they may have reinforced existing asymmetries in hospital attractiveness, further differentiating the latent positions of ASLs in the post-pandemic years.

A comparison of the latent geometries recovered by \texttt{Mult\_Geo\_Offset} and \texttt{Base\_Geo\_Offset} shows that the two models produce broadly consistent representations, with Procrustes correlations ranging from 0.93 in 2018 to 0.97 in 2024. The two specifications therefore recover a similar picture of the underlying mobility network, with the residual divergence reflecting the fact that in \texttt{Base\_Geo\_Offset} directional asymmetries are absorbed by the latent space, whereas in \texttt{Mult\_Geo\_Offset} they are captured explicitly by the sender and receiver parameters $\gamma_i$ and $\theta_j$.

\subsection{Directional heterogeneity in mobility} 

\begin{figure}[htbp]
\centering
\includegraphics[width=0.95\textwidth]{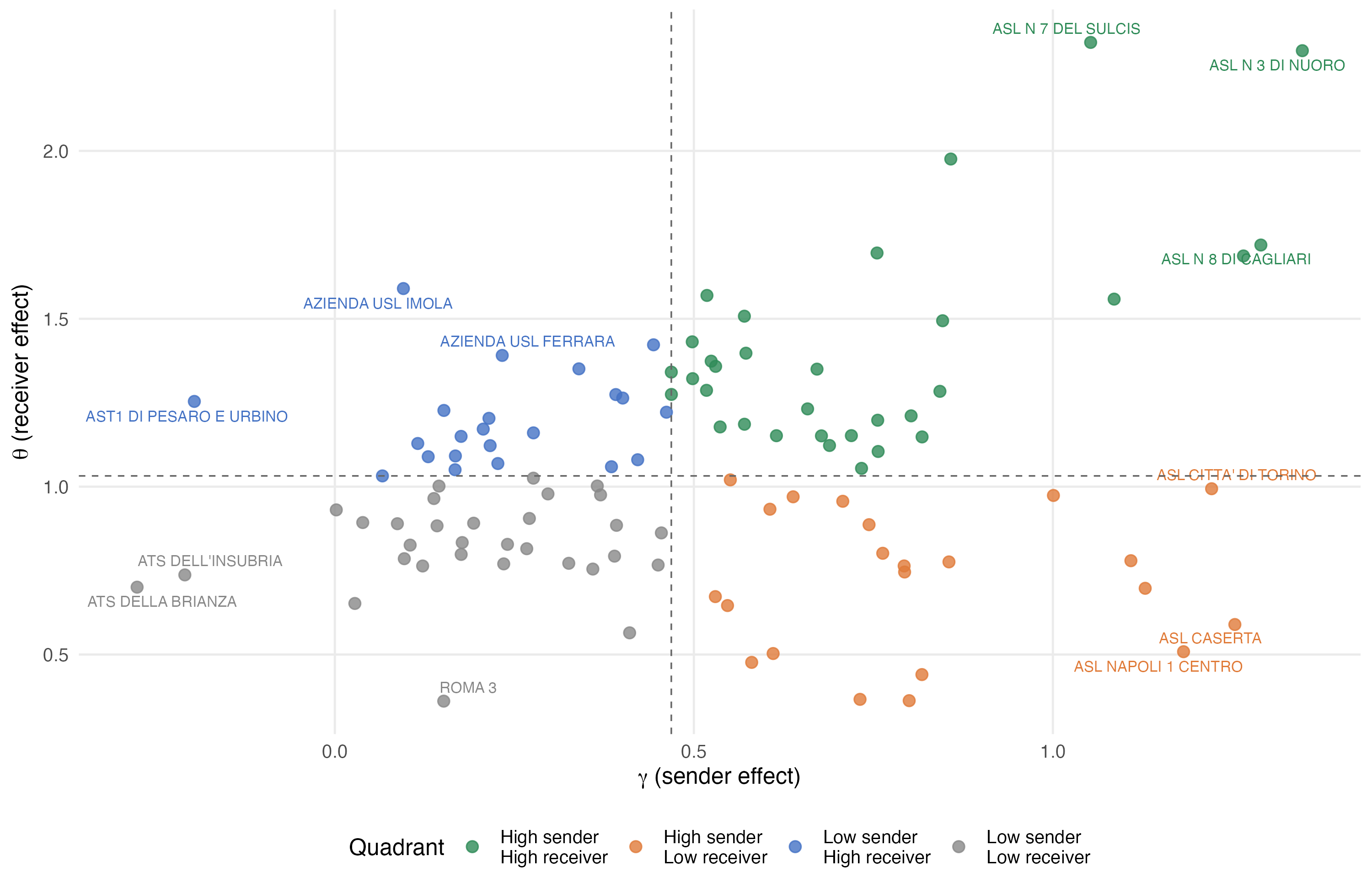}
\caption{Posterior mean sender effects $\gamma_i$ and 
receiver effects $\theta_j$ for all 109 ASLs under 
\texttt{Mult\_Geo\_Offset}. Both effects are measured relative to reference nodes fixed at the sample median of out-degree and in-degree respectively. Dashed lines are drawn at the sample medians of the estimated effects. ASLs are classified into four quadrants accordingly, and the three most extreme nodes in each quadrant are labelled.}
\label{fig:sender_receiver}
\end{figure}

Figure~\ref{fig:sender_receiver} displays the estimated posterior means of the sender effects $\gamma_i$ and receiver effects $\theta_j$ for all 109 ASLs under \texttt{Mult\_Geo\_Offset}. Both effects are identified relative to reference nodes chosen at the sample median of out-degree and in-degree respectively, and are estimated net of the exposure term $\log E_{ij}^{(t)}$.
They therefore measure outward propensity and attractiveness beyond what is predicted by territorial size alone.

Under this specification, Sardinian ASLs dominate the upper-right quadrant, combining high sender and high receiver effects. 
This configuration reflects the layered structure of mobility within the island: a limited set of hub ASLs both pull in patients from nearby, smaller areas and generate outflows towards other island hubs or the mainland. 
These roles remain undetected in specifications without exposure correction, since Sardinian ASLs are too small to stand out in terms of absolute flow volumes.
The high sender/low receiver quadrant is occupied primarily by the Campania cluster: all ASLs display high outward propensity net of size, consistent with the well-documented passive mobility from Campania towards other regions \citep{carnazza2025monetary, balia2018interregional}. 
Several Piemonte ASLs also exhibit high sender effects, including ASL Città di Torino ($\gamma = 1.221$), CN2 ($\gamma = 1.109$), and CN1 ($\gamma = 1.001$). However, analysis of the observed flow matrices reveals that these outflows are primarily directed towards other Piemonte ASLs, rather than indicating inter-regional passive mobility.
At the lower end of the sender distribution, ATS della Brianza ($\gamma = -0.276$), ATS dell'Insubria ($\gamma = -0.209$) and ATS della Città Metropolitana di Milano ($\gamma = 0.002$) send fewer patients than the reference node net of their large size, reflecting the strong self-sufficiency of the Lombardia hub system. 
The Autonomous Provinces of Bolzano ($\gamma = 0.1758$) and Trento ($\gamma = 0.1759$) similarly display low sender effects, consistent with their well-documented capacity to retain patients locally \citep{EuroObsItaly2024}. 

The comparison with \texttt{Mult\_Geo} demonstrates how strongly the interpretation of sender and receiver effects depends on the inclusion of the exposure term.
Without offset correction, the top receivers are the large northern facilities: ATS della Città Metropolitana di Milano ($\theta = 2.506$), Azienda USL Bologna ($\theta = 2.313$), ASL Città di Torino ($\theta = 2.098$) and Roma 1 ($\theta = 2.068$). 
These rankings reflect absolute flow volumes --- large ASLs attract many patients simply because of their size --- rather than structural attractiveness. 
Once the exposure term absorbs size effects, this ranking is entirely reversed in favour of the Sardinian and Sicilian hubs previously described. 
The Pearson correlation between posterior means of $\theta_j$ across the two specifications is essentially negligible at 0.03, indicating that the two models capture different aspects of receiver heterogeneity; sender effects are more stable across specifications (correlation 0.58).
Detailed results for \texttt{Mult\_Geo} are reported in the Appendix.

\subsection{Overdispersion across years}

Table~\ref{tab:r_summary} reports the posterior means 
and 95\% credible intervals for the negative binomial 
dispersion parameter $r^{(t)}$ under \texttt{Mult\_Geo\_Offset}. 
All estimates fall well below unity, confirming the strong overdispersion already apparent in Table~\ref{tab:structure}.

\begin{table}[htbp]
\centering
\caption{Posterior summaries of the dispersion parameter 
$r^{(t)}$ under \texttt{Mult\_Geo\_Offset} (2018--2024).}
\label{tab:r_summary}
\begin{tabular}{lrrr}
\toprule
Year & Mean & 2.5\% & 97.5\% \\
\midrule
2018 & 0.481 & 0.397 & 0.563 \\
2019 & 0.601 & 0.515 & 0.689 \\
2020 & 0.615 & 0.526 & 0.703 \\
2021 & 0.644 & 0.558 & 0.736 \\
2022 & 0.539 & 0.462 & 0.621 \\
2023 & 0.534 & 0.464 & 0.607 \\
2024 & 0.480 & 0.419 & 0.547 \\
\bottomrule
\end{tabular}
\end{table}

The temporal evolution of $r^{(t)}$ shows an increase from 0.481 in 2018 to a peak of 0.644 in 2021, suggesting a temporary reduction in overdispersion during the pandemic year.
This is consistent with the sharp contraction of elective surgical activity reported in Table~\ref{tab:admissions}. As the overall volume of hip replacement procedures declined and mobility flows became more concentrated along established corridors, the variability of the counts decreased relative to their mean. After the peak in 2021, $r^{(t)}$ declines steadily through 2022--2024, reaching 0.480 by the end of the study period.
This decrease points to a renewed increase in heterogeneity in flow intensities during the post-pandemic phase, as some mobility corridors expanded markedly while others remained relatively stable, leading to a wider dispersion of counts.

Under \texttt{Base\_Geo\_Offset}, the estimated values of $r^{(t)}$ are consistently lower, ranging from 0.356 to 0.501, and display a different temporal pattern: rather than peaking in 2021, they peak earlier in 2020 and decline more sharply through 2022--2024.
This difference reflects the role of the sender and receiver parameters in \texttt{Mult\_Geo\_Offset}: by explicitly capturing directional heterogeneity through $\gamma_i$ and $\theta_j$, this specification absorbs part of the variability that \texttt{Base\_Geo\_Offset} leaves to the dispersion parameter, resulting in higher and more stable estimates of $r^{(t)}$. The consistently low values of $r^{(t)}$ across all years and both specifications confirm that overdispersion is a structural feature of these data rather than an artefact of model choice.

\section{Discussion}
\label{sec:discussion}

In this work we proposed a dynamic latent space model for weighted networks with a hurdle negative binomial likelihood, and applied it to study inter-ASL hip replacement mobility in Italy over the period 2018--2024. The framework jointly addresses three structural features of healthcare mobility data: excess zeros, strong overdispersion, and network dependence among dyads. By embedding local health authorities (Aziende Sanitarie Locali, ASL) within a low-dimensional latent space with time-varying positions, the model yields an interpretable geometric representation of the mobility network and its evolution, while the multiplicative sender--receiver component captures directional asymmetries in outward propensity and ASL attractiveness net of territorial size.

The simulation study confirmed that the proposed MCMC algorithm recovers the latent geometry with high accuracy across a broad range of scenarios, including high sparsity and strong overdispersion. The sender and receiver effects are estimated with high accuracy, with correlations between posterior means and true values consistently exceeding 0.98. 
A limitation identified in the simulation concerns the intercept parameter $\alpha^{(t)}$, which displays a small positive bias under high overdispersion, particularly in the multiplicative specification. This bias does not compromise the recovery of the latent structure, which constitutes the primary inferential target of the analysis.

The application to Italian hip replacement data revealed a persistent geographical structure in the latent space, with clusters reflecting the regional referral networks. 
Southern ASLs tend to occupy more peripheral positions associated with high outward mobility and limited local retention capacity. The island regions of Sicilia and Sardegna form distinct latent clusters that reflect their specific logistical barriers and intra-island referral patterns. These structural features remained stable throughout the observation period, indicating that the north--south gradient in Italian healthcare is a persistent feature of the system rather than a transient configuration.

The sender and receiver effects estimated under \texttt{Mult\_Geo\_Offset} reveal a pattern that differs markedly from what would be obtained without the exposure correction. 
Once size effects are absorbed by the exposure term, several Sardinian and Sicilian hub ASLs appear among the top nodes by both sender and receiver effects, rather than the large northern facilities that dominate absolute flow volumes. 
These ASLs simultaneously attract patients from surrounding smaller territories and generate outflows towards other island hubs or the mainland. 
By contrast, various Emilia-Romagna ASLs display relatively low sender effects and moderate-to-high receiver effects net of size, consistent with a system that retains patients locally while attracting flows from neighbouring territories. 
The Campania cluster dominates the high sender/low receiver quadrant in both specifications, partially confirming the documented passive mobility from southern Italy \citep{carnazza2025monetary, balia2018interregional}, while the strong self-sufficiency of Lombardia ASLs is consistently captured by low or negative sender effects net of their large size.

The temporal evolution of the latent space captures the disruption induced by the COVID-19 pandemic and the subsequent reorganisation of elective surgical activity. 
The increase in latent dispersion between 2019 and 2021 suggests a spatially uneven contraction and recovery of hip replacement volumes, while the continued increase through 2022--2024 is consistent with the recovery across regions, linked to the gradual implementation of post-pandemic investments under the PNRR.
The estimated dispersion parameter $r^{(t)}$ provides complementary evidence: its temporary increase in 2020--2021 indicates a concentration of flows along established mobility corridors during the pandemic, whereas the subsequent decline suggests increasing heterogeneity in flow intensities as elective activity resumed unevenly across regions.

The comparison between \texttt{Mult\_Geo\_Offset} and \texttt{Base\_Geo\_Offset} highlights a key difference in what the two specifications capture.
In \texttt{Base\_Geo\_Offset}, sender and receiver effects are absent and directional heterogeneity is absorbed by the latent space.
By contrast, \texttt{Mult\_Geo\_Offset} assigns this heterogeneity to the explicit parameters $\gamma_i$ and $\theta_j$, leaving the latent space to represent the residual symmetric structure of patient exchanges.
The Procrustes correlations between the latent configurations of the two models range from 0.93 to 0.97 across years, indicating that the underlying geometry remains stable. 

A comparison between \texttt{Mult\_Geo\_Offset} and \texttt{Mult\_Geo} further shows how the inclusion of the exposure term changes the interpretation of the receiver effects.
Without the correction, $\theta_j$ captures absolute flow volumes and favours large northern facilities.
When the offset is included, $\theta_j$ instead measures structural attractiveness net of territorial size and ASL capacity, highlighting the role of smaller hub ASLs.

The near-zero correlation between the receiver effects in the two specifications (0.03) confirms that the difference reflects a substantive change in the quantity being estimated. 
The choice between the two formulations should depend on whether the objective is to analyse absolute attractiveness or attractiveness adjusted for territorial exposure.

The proposed framework and the accompanying \texttt{R} and \texttt{C++} 
implementation\footnote{\url{https://github.com/ceciliamanente/countLSM}} 
-- provide a flexible tool for the statistical analysis of dynamic healthcare mobility networks.
Beyond the Italian case, the methodology is directly applicable to any setting where directed weighted flows among territorial units are observed over time.

\section*{Data availability}
The data used in this analysis are administrative hospital discharge records held by the Italian Ministry of Health and the National Agency for Regional Health Services (AGENAS). Access is subject to institutional data agreements.

\bibliographystyle{plainnat}

\bibliography{reference}

\appendix

\section{Coverage of posterior credible intervals}
\label{app:sim_coverage}

\begin{figure}[htbp]
\centering
\includegraphics[width=1\textwidth]{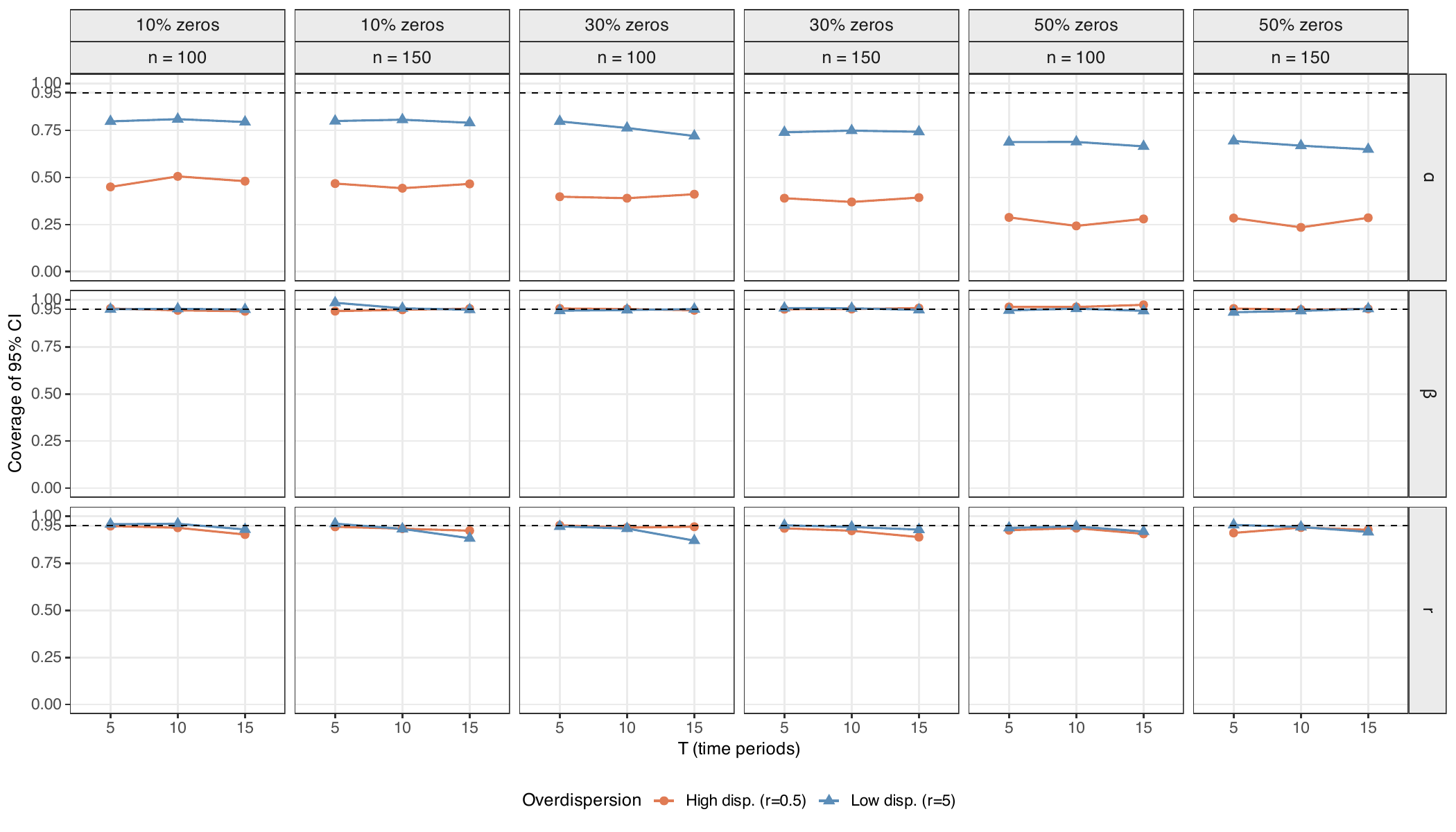}
\caption{Coverage of 95\% credible intervals for $\alpha^{(t)}$, $\beta^{(t)}$, and $r^{(t)}$ under the baseline model, across all combinations of network size $n$, panel length $T$, sparsity level, and overdispersion regime. The dashed horizontal line marks the nominal 0.95 level.}
\label{fig:coverage_base_simulation}
\end{figure}

\begin{figure}[htbp]
\centering
\includegraphics[width=1\textwidth]{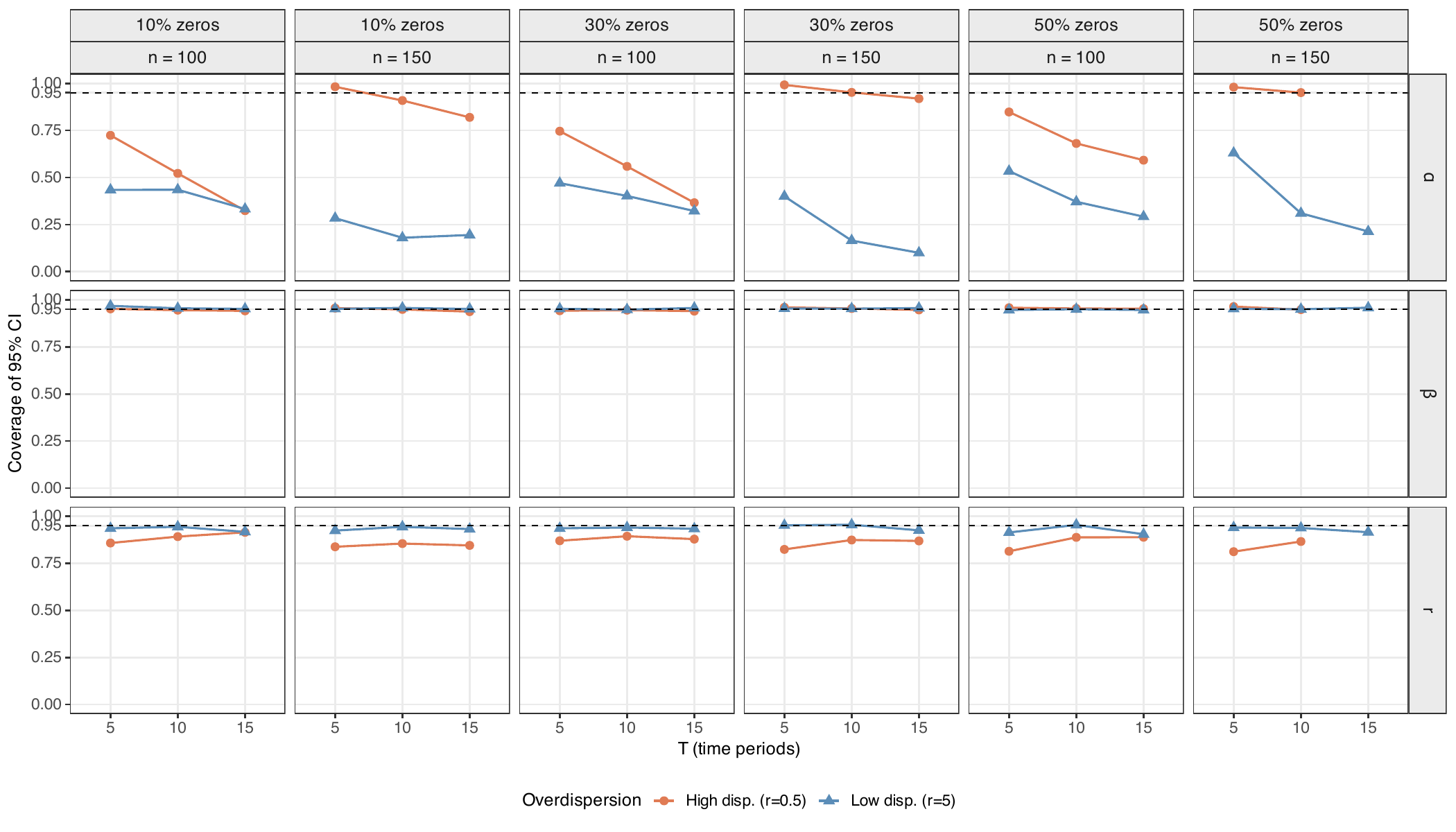}
\caption{Coverage of 95\% credible intervals for $\alpha^{(t)}$, $\beta^{(t)}$, and $r^{(t)}$ under the multiplicative model, across all combinations of network size $n$, panel length $T$, sparsity level, and overdispersion regime. The dashed horizontal line marks the nominal 0.95 level.}
\label{fig:coverage_molt_simulation}
\end{figure}

Figures~\ref{fig:coverage_base_simulation} 
and~\ref{fig:coverage_molt_simulation} report the empirical coverage of the 95\% credible intervals for $\alpha^{(t)}$, $\beta^{(t)}$, and $r^{(t)}$ under the baseline and multiplicative specifications, respectively. 
Coverage for $\beta^{(t)}$ and $r^{(t)}$ is close to the nominal 0.95 level across all settings. 
Coverage for $\alpha^{(t)}$ is substantially below the 
nominal level under high overdispersion in both models, 
consistent with the positive bias discussed in Section 4.2 of the paper.

\section{Baseline model with exposure adjustment and geographical distance}
\label{app:base_geo_offset}

This appendix reports the full results for \texttt{Base\_Geo\_Offset}, the main alternative to the preferred specification discussed in Section 5.
This model includes the exposure term and geographical distance in the hurdle component, but does not include sender and receiver effects. Directional heterogeneity is therefore absorbed entirely by the latent space rather than delegated to explicit node-level parameters.

\subsection{Latent geometry}
\begin{figure}[htbp]
\centering
\includegraphics[width=1\textwidth]{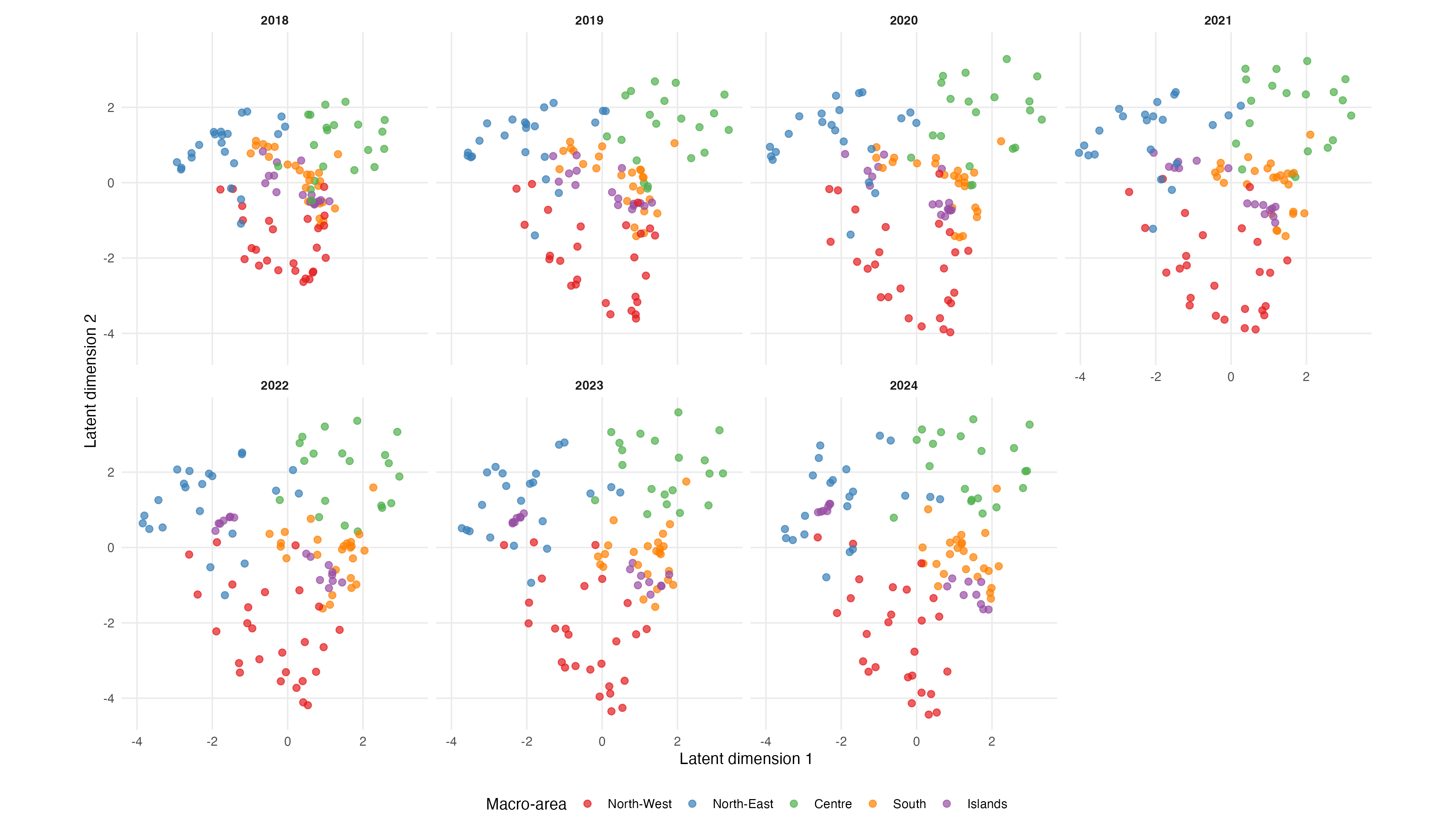}
\caption{Temporal evolution of latent positions for 109 Italian ASLs (2018--2024) under \texttt{Base\_Geo\_Offset}, coloured by macro-area.}
\label{fig:latent_macroarea_bgo}
\end{figure}

\begin{figure}[htbp]
\centering
\includegraphics[width=1\textwidth]{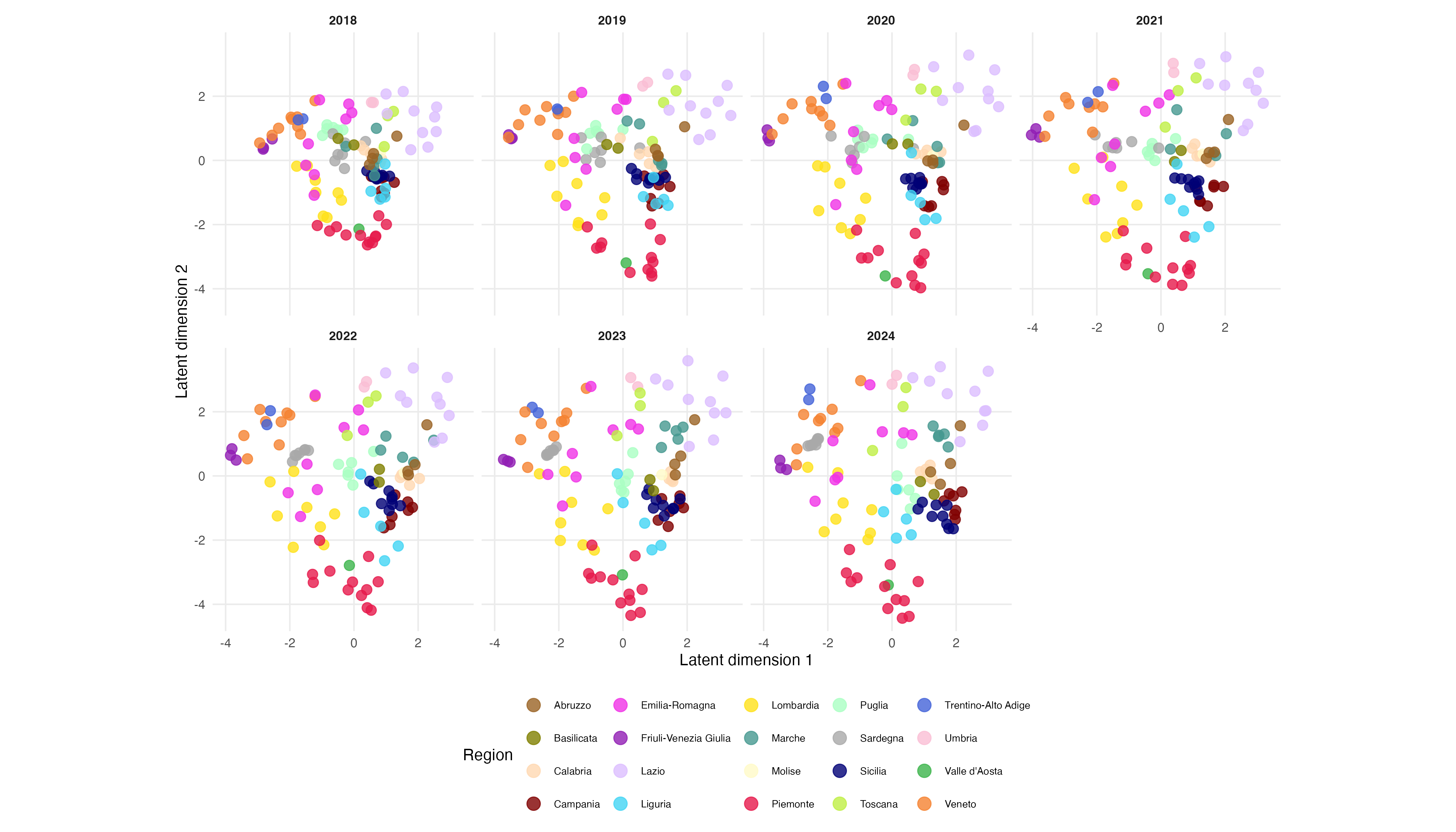}
\caption{Temporal evolution of latent positions for 109 Italian ASLs (2018--2024) under \texttt{Base\_Geo\_Offset}, coloured by region.}
\label{fig:latent_region_bgo}
\end{figure}

Figures~\ref{fig:latent_macroarea_bgo} and~\ref{fig:latent_region_bgo} show the posterior mean latent positions coloured by macro-area and region respectively.
The broad geographical clustering visible under \texttt{Mult\_Geo\_Offset} is reproduced here: ASLs from the same macro-area tend to occupy nearby positions, and the two island clusters remain distinct throughout the observation period. 
However, the latent configuration is markedly more compressed: the mean Euclidean distance from the centroid increases from 1.55 in 2018 to 2.34 in 2024, compared with 2.06 to 3.16 under \texttt{Mult\_Geo\_Offset}. 

\subsection{Overdispersion}

Table~\ref{tab:r_bgo} reports the posterior summaries of the dispersion parameter $r^{(t)}$ under \texttt{Base\_Geo\_Offset}. 
Compared with \texttt{Mult\_Geo\_Offset}, the estimated values are consistently lower across all years, ranging from 0.356 to 0.501, and peak earlier in 2020 rather than 2021. 
As discussed in Section 5, this difference reflects the role of the sender and receiver parameters: by absorbing directional heterogeneity explicitly, \texttt{Mult\_Geo\_Offset} leaves less residual variability to be captured by the dispersion parameter.

\begin{table}[htbp]
\centering
\caption{Posterior summaries of the dispersion parameter $r^{(t)}$ under \texttt{Base\_Geo\_Offset} (2018--2024).}
\label{tab:r_bgo}
\begin{tabular}{lrrr}
\toprule
Year & Mean & 2.5\% & 97.5\% \\
\midrule
2018 & 0.451 & 0.379 & 0.526 \\
2019 & 0.468 & 0.401 & 0.540 \\
2020 & 0.501 & 0.420 & 0.585 \\
2021 & 0.457 & 0.384 & 0.535 \\
2022 & 0.408 & 0.346 & 0.474 \\
2023 & 0.411 & 0.352 & 0.475 \\
2024 & 0.356 & 0.299 & 0.417 \\
\bottomrule
\end{tabular}
\end{table}

\section{Multiplicative S/R model with geographical distance}
\label{app:mult_geo}

This appendix reports the full results for \texttt{Mult\_Geo}, the specification including sender and receiver effects and geographical distance in the hurdle component but without any exposure term.
In this specification $\gamma_i$ and $\theta_j$ capture absolute outward propensity and attractiveness rather than effects net to territorial size.

\subsection{Latent geometry}
\begin{figure}[htbp]
\centering
\includegraphics[width=1\textwidth]{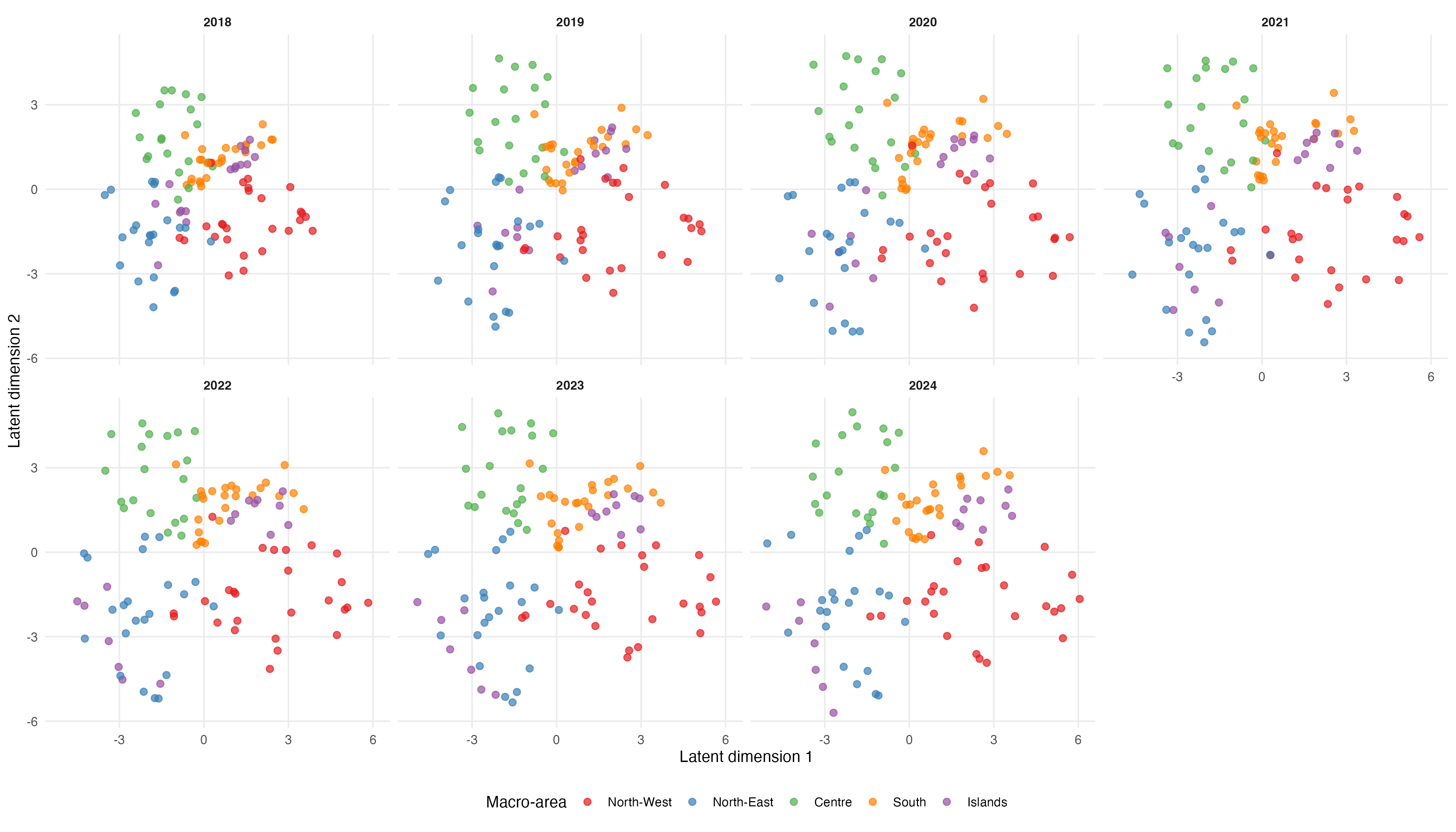}
\caption{Temporal evolution of latent positions for 109 Italian ASLs (2018--2024) under \texttt{Mult\_Geo}, coloured by macro-area.}
\label{fig:latent_macroarea_mg}
\end{figure}

\begin{figure}[htbp]
\centering
\includegraphics[width=0.95\textwidth]{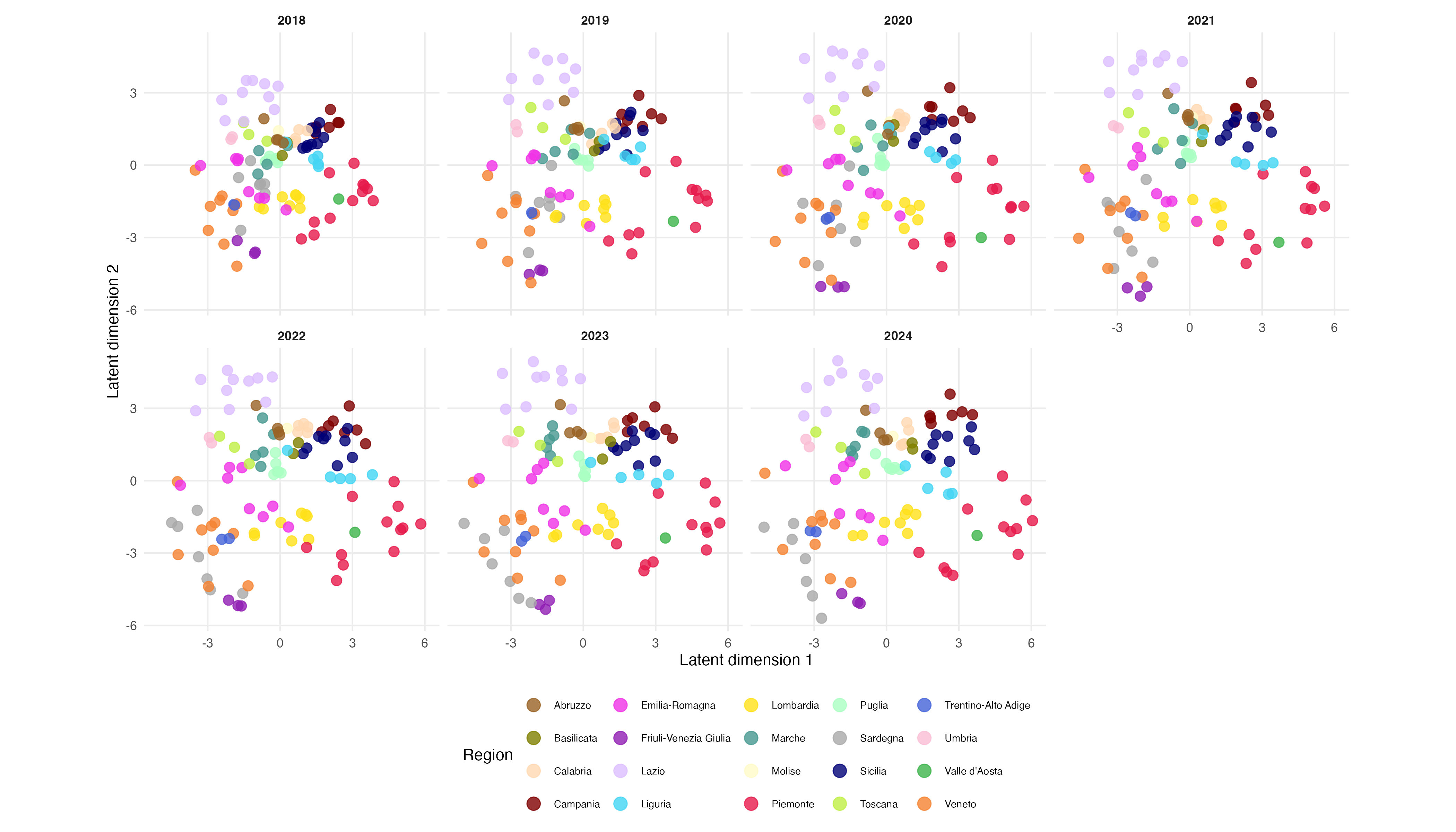}
\caption{Temporal evolution of latent positions for 109 Italian ASLs (2018--2024) under \texttt{Mult\_Geo}, coloured by region.}
\label{fig:latent_region_mg}
\end{figure}

Figures~\ref{fig:latent_macroarea_mg} and~\ref{fig:latent_region_mg} show the posterior mean latent positions coloured by macro-area and region respectively.
The broad geographical clustering visible under \texttt{Mult\_Geo\_Offset} is reproduced here: ASLs from the same macro-area tend to occupy nearby positions, and the two island clusters remain distinct throughout the observation period. 
The dispersion of latent positions increases from 2.13 in 2018 to 3.25 in 2024: a pattern close to that of the preferred specification.
This is consistent with the high Procrustes correlation between the two models ($>$0.996 in every year), indicating that the latent geometries are essentially identical.
The inclusion of the exposure term therefore does not alter the recovered network structure but changes the interpretation of the sender and receiver effects.

\subsection{Sender and receiver effects}

\begin{figure}[htbp]
\centering
\includegraphics[width=1\textwidth]{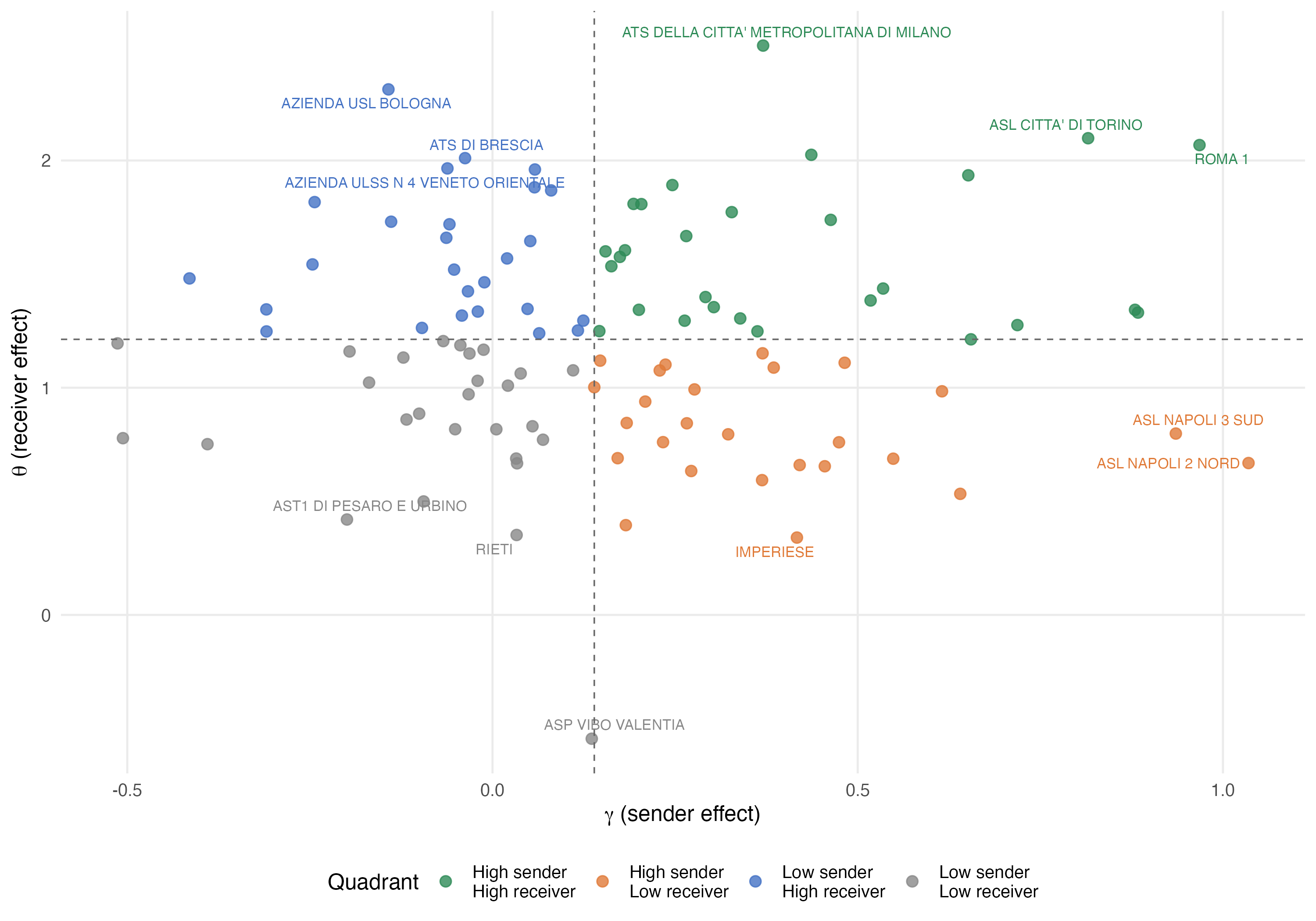}
\caption{Posterior mean sender effects $\gamma_i$ and receiver effects $\theta_j$ for all 109 ASLs under \texttt{Mult\_Geo}. Dashed lines are drawn at the sample medians of the estimated effects. ASLs are classified into four quadrants accordingly, and the three most extreme nodes in each quadrant are labelled.}
\label{fig:sender_receiver_mg}
\end{figure}

Figure~\ref{fig:sender_receiver_mg} displays the posterior mean sender and receiver effects. 
Without exposure correction, the top receivers are the large northern ASLs that dominate absolute flow volumes: ATS della Città Metropolitana di Milano ($\theta = 2.506$), Azienda USL Bologna ($\theta = 2.313$), ASL Città di Torino ($\theta = 2.098$), Roma 1 ($\theta = 2.068$) and Azienda ULSS N 2 Marca Trevigiana ($\theta = 2.025$). 
At the opposite end, ASP Vibo Valentia is the only ASL with a negative receiver effect ($\theta = -0.545$), indicating minimal attractiveness even in absolute terms.
The top senders are dominated by the Campania cluster: ASL Napoli 2 Nord ($\gamma = 1.035$), ASL Napoli 3 Sud ($\gamma = 0.935$), ASL Caserta ($\gamma = 0.884$) and ASL Napoli 1 Centro ($\gamma = 0.880$).
Roma 1 appears in both the top senders and top receivers, occupying the upper-right quadrant and reflecting its dual role as a major generator and attractor of patient flows in absolute terms. 
At the lower end of the sender distribution, the Autonomous Provinces of Bolzano ($\gamma = -0.513$) and Trento ($\gamma = -0.415$) display strongly negative sender effects, consistent with their self-sufficiency.

\end{document}